\documentclass[12pt, preprint]{aastex}

\usepackage{lscape}
\usepackage{indentfirst}
\usepackage{graphicx}
\usepackage[varg]{txfonts}

\begin{document}

\begin{center}
\textbf{\Large From planetesimals to terrestrial planets: $N$-body simulations including the effects of nebular gas and giant planets} 

Ryuji Morishima$^{1,2}$, Joachim Stadel$^{3}$,  Ben Moore$^{3}$

Morishima@lasp.colorado.edu

1: Jet Propulsion Laboratory/California Institute of Technology, Pasadena, CA, USA

2: Institute of Geophysics and Planetary Physics/UCLA, Los Angels, CA, USA

3: Institute for Theoretical Physics, University of Zurich, Switzerland


\end{center}

\vspace{8em}
Accepted for publication in Icarus 

\vspace{3em}

Total manuscript page :  51

Table: 2

Figures: 19

\clearpage

\hspace{-2em}\textbf{Proposed Running Head:}
 
\textbf{Formation of terrestrial planets with gas and giant planets} 

\vspace{2em}

\hspace{-2em}\textbf{Editorial correspondence to:}

\hspace{-2em}Ryuji Morishima

\begin{description}
\item 
Jet Propulsion Laboratory, M/S 230-205, 4800 Oak Grove Drive, Pasadena, CA 91109, USA

\item Phone:+1 818 393 1014; Fax:+1 818 393 4495

\item e-mail: ryuji.morishima@jpl.nasa.gov

\end{description}

\clearpage

\begin{abstract}
{
We present results from a suite of $N$-body simulations that follow the formation and accretion
history of the terrestrial planets using a new parallel treecode that we have developed. 
We initially place 2000 equal size planetesimals
between 0.5--4.0 AU and the collisional growth is followed until
the completion of planetary accretion ($>$ 100 Myr). 
A total of 64 simulations were carried out to explore sensitivity to
the key parameters and initial conditions.
All the important effect of gas in laminar disks are taken into account:
the aerodynamic gas drag, the disk-planet interaction including Type I migration,
and the global disk potential which causes
inward migration of secular resonances as the gas dissipates.
We vary the initial total mass and spatial distribution of the planetesimals, 
the time scale of dissipation of nebular gas (which dissipates uniformly in space and exponentially in time),
and orbits of Jupiter and Saturn.  
We end up with one to five planets in the terrestrial region. 
In order to maintain sufficient mass in this region in the presence of Type I migration, 
the time scale of gas dissipation needs to be 1-2 Myr.
The final configurations and collisional histories strongly depend on the orbital eccentricity of Jupiter.
If today's eccentricity of Jupiter is used, then most of bodies in the asteroidal region are 
swept up within the terrestrial region owing to the inward migration of the secular resonance, 
and giant impacts between protoplanets occur most commonly around 10 Myr.   
If the orbital eccentricity of Jupiter is close to zero, as 
suggested in the Nice model, the effect of the secular resonance is negligible 
and a large amount of mass stays for a long period of time 
in the asteroidal region.  With a circular orbit for Jupiter,  
giant impacts usually occur around 100 Myr, consistent with the accretion time scale indicated from isotope records.
However, we inevitably have an Earth size planet at around 2 AU in this case.
It is very difficult to obtain spatially concentrated terrestrial planets together 
with very late giant impacts, as long as we include all the above 
effects of gas and assume initial disks similar to the minimum mass solar nebular. 

}

\end{abstract}

\section{Introduction}
Accretion models for the formation of the terrestrial planets need to reproduce the following properties of our solar system.  
(1) The spatial mass distribution of the terrestrial planets is radially concentrated. Namely,
the Earth and Venus contain most of the mass inside of Jupiter's orbit, whereas 
the masses of Mars and Mercury are an order of magnitude smaller than the Earth's mass ($M_{\oplus}$) 
and the total mass of asteroids is only $5\times 10^{-4}$ $M_{\oplus}$. 
The orbital separation between Earth and Venus is only 0.3 AU. 
(2) The orbits of the Earth and Venus are nearly circular and coplanar even though orbital excitements are expected in 
the late stage of accretion where large protoplanets experience mutual collisions. 
This probably indicates that there were some damping mechanisms to reduce
their orbital eccentricities and inclinations near the end of their formation processes.
(3) Isotope records such as W/Hf and U/Pb for the Earth and Moon 
indicate that the Moon-forming giant impact occurred $\sim 100$ Myr after the beginning 
of the solar system (Touboul et al., 2007; All\'{e}gre et al., 2008). 
Some accretion models can reproduce two of the three properties but 
there is no accretion model that satisfies all three properties. 

$N$-body simulations that follow only the gravitational interactions of protoplanets (or with the giant planets) usually produce too large orbital eccentricities 
of the final planets (Chambers, 1998; Agnor et al., 1999; Raymond et al., 2004; Kokubo et al., 2006). To resolve this issue two damping mechanisms have been proposed:
one is by remnant planetesimals (Chambers, 2001; O'Brien et al., 2006; Raymond et al., 2006, 2009; Morishima 
et al., 2008) and another is by remnant nebular gas (Agnor and Ward, 2002; Kominami et al., 2002, 2004; 
Nagasawa et al., 2005; Ogihara et al., 2007; Thommes et al., 2008). 

In particular, the dynamical shake-up model of Nagasawa et al. (2005) and  Thommes et al. (2008)
is the most successful model so far in reproducing both the small orbital eccentricities and the radial mass concentration. 
Their model initially places a few tens of protoplanets embedded in a gas disk and takes into account the effect of 
the gas giant planets with orbital eccentricities comparable to or slightly larger than the present values 
(the present eccentricities of both Jupiter and Saturn are $\sim$ 0.05).
The secular resonance with Jupiter migrates from $\sim$4.0 to 0.6 AU with dissipation of the gas disk  and sweeps protoplanets 
toward the current location of the Earth. Mutual impacts between protoplanets occur even when a large 
amount of gas remains, and the remnant gas disk reduces the orbital eccentricities of formed Earth-like planets. 

If the formation of the Moon occurred very late, as indicated by recent isotope measurements 
(Touboul et al., 2007), significant damping due to remnant gas may be problematic for the following reasons.
Significant depletion of gas already at $\sim$ 10 Myr is suggested from direct measurements of line emissions of circumstellar disks 
(Pascucci et al. 2006) and from H$\alpha$ emission profiles of central stars due to accretion shock (Muzerolle et al., 2000), 
although we cannot rule out the possibility that our protosolar nebular was exceptionally long-lived.    
Another potential problem for long-lived gas disks is depletion of the solid component due to Type I migration, which is neglected in 
the model of Nagasawa et al. (2005) and  Thommes et al. (2008).
Once a protoplanet becomes as massive as Mars in a gas disk,
the gravitational torque from spiral density waves launched by the protoplanet 
causes rapid inward migration of the protoplanet, referred to as Type I migration
(Goldreich and Tremaine, 1980; Ward, 1986, 1997; Tanaka et al., 2002).
McNeil et al. (2005) and Daisaka et al. (2006) conducted $N$-body simulations taking Type I migration into account 
and found that the decay time scale for a gas disk needs to be $\sim 1$ Myr 
in order to retain sufficient solid mass for the formation of Earth-size planets.

Some remnant planetesimals are likely to have existed even at 100 Myr particularly in the asteroidal region. \footnote[1]{
In the present paper, we loosely refer to the outside and inside of 2 AU as 
the asteroidal and terrestrial regions, respectively.}   
Several authors have conducted $N$-body simulations starting with protoplanets and remnant planetesimals
(Chambers, 2001; Chambers and Cassen, 2002; O'Brien et al. 2006; Raymond et al. 2006, 2009). 
Their results show that final eccentricities of planets tend to be lower with increasing total mass of remnant planetesimals.
Even with the same total mass of planetesimals, the damping is more effective 
when the size of an individual planetesimal is smaller and planetesimals are placed uniformly over both the 
terrestrial and asteroidal regions rather than in the asteroidal region only (O'Brien et al., 2006; Raymond et al., 2006). 
An eccentricity of Jupiter comparable to or slightly larger than the present value is preferred in regard to 
the radial mass concentration (Chambers and Cassen, 2002; O'Brien et al., 2006; Raymond et al., 2009),  
although none of simulations have succeeded in reproducing the very high radial mass concentration of the current terrestrial planets.
The spatial and size distributions of remnant planetesimals 
are arbitrarily assumed in these simulations. In order to clarify these distributions  
one needs to follow the growth of planetesimals and protoplanets in the runaway growth stage. 
In addition, the above simulations neglect the effects of gas. The protosolar nebular must have existed at least until the formation 
of Jupiter, and probably survived somewhat longer. Even a rapid dissipation of gas of the order of 0.1 Myr in the presence of Jupiter 
strongly affects the orbital evolution of asteroidal bodies (Nagasawa et al., 2000).

In the present study we perform $N$-body simulations starting with a wide planetesimal disk and continue 
until the end of the accretion phase. Since we directly calculate the runaway growth of the protoplanets, we can obtain reasonable
spatial and size distributions of remnant planetesimals in the late stage. 
We take into account the gas giant planets and all the possible effects of gas in laminar disks: aerodynamic gas drag, disk-planet 
interaction including Type I migration, and global disk potential. In a previous study, Morishima et al. (2008) also conducted $N$-body simulations that followed the evolution of a protoplanetary disk starting with planetesimals only until the end of accretion. However, their simulations started from radially narrow annuli
(widths of 0.3-0.5 AU) and neglected any external forces.
In Sec.~2,  we introduce our new $N$-body code, review the effects of nebular gas, and outline our choice of simulation parameters.
In Sec.~3, we show time evolution of two contrasting examples, present the results of all simulations and compare them with the solar system.
In principle, none of our simulations succeed in satisfying all the three constraints listed in the first paragraph of this section.  
In Sec.~4, we discuss possible solutions for this problem. In Sec.~5, we give our conclusions.  

\section{Method}

\subsection{$N$-body code}
Any $N$-body code consists of a gravity solver to calculate  the mutual
gravity force and an integrator to update positions and velocities for a next time step.  
Existing $N$-body codes applied for the runaway growth stage have 
good gravity solvers, such as tree methods (Richardson et al., 2000; Stadel, 2001) or 
special hardware such as HARP or GRAPE (Kokubo and Ida, 1996).  
On the other hand, $N$-body codes applied for the late stage have good integrators, such as 
SyMBA (Duncan et al., 1998) and Mercury (Chambers, 1999), 
so that one can take a large time step size, while solving the mutual gravity by direct $N^2$ calculation.
The new $N$-body code we have developed has both a good gravity solver and a good integrator, 
allowing us to follow accretion stages of planetary systems with large $N$ and over a large number of orbits. 
The code is still being upgraded and details of the complete version will be reported elsewhere. 
Here we briefly describe the version of the code used for simulations in this paper. 

The gravity solver we use is a parallel-tree method, PKDGRAV (Stadel, 2001).
The original version of PKDGRAV uses a leap-frog integration scheme, where individual particles
are updated in hierarchical time step blocks, each block with a factor of two smaller time-step according 
to the local dynamical time of its particles.
Richardson et al. (2000) added some functions for planetary accretion to the code and conducted 
simulations for the runaway growth stage. 
In the tree method, all particles are divided into hierarchal tree-structured groups.
Then, the gravity forces from distant particle groups are calculated using their multipole expansions.
The zeroth-order multipole expansion corresponds to the gravity from the mass center of a particle group. 
In order to keep good accuracy, we include multipole expansions up to the fourth order.     
The computer time of gravity calculation using this gravity solver is only proportional to 
$N\log{N}$. We find a significant speed up in the gravity calculation with our tree method 
as compared with the direct $N^2$ calculation when $N > 200$.
In addition, the tree method is suitable for parallel computing, and this results in further speed up when $N > 1000$.
Since the number of particles decreases with growth of planets, we switch calculation methods from parallel-tree to 
serial-tree, and to serial-direct $N^2$ calculations.  
The error in the total energy due to the tree method appears as a random walk, and thus is proportional to the square root 
of the number of time steps. Therefore, with some short additional simulations, we can estimate the accumulated error after 
a long term evolution of a system. The accuracy of the method is controlled by the opening angle for tree cells and 
we set it to be 0.5 in the present paper.  
We confirm that the expected error due to the gravity calculation is sufficiently smaller than 
the error due to the integrator's, which rises most notably during close encounters.
   
The integrator implemented in the code 
is a mixed-variable symplectic (MVS) integrator, SyMBA (Duncan et al., 1998). 
In MVS integrators such as SyMBA and Mercury, the Hamiltonian is split into
a part which describes the Keplerian motion around the central star and another part which describes 
the perturbation from other bodies. 
The former part corresponds to updates of positions and velocities along Kepler orbits. 
This can be done nearly analytically.
The latter part corresponds to velocity changes due to the perturbation from other bodies, which we calculate using  
the tree method described above.
These types of integrators are very accurate and allow us to take very large time steps
for long term simulations.  We use a time step size, $\Delta t$, of 6 days. 
This is comparable to those used in previous high-resolution 
$N$-body simulations using MVS integrators (O'Brien et al., 2006; Raymond et al., 2006).
Practically, the time step size needs to be chosen so that a fraction of particles in close encounters defined below 
is small enough.

In close encounter search, we first search for 10--20 nearest neighbors
of each particle using the same algorithm in the tree SPH code, Gasoline (Wadsley et al., 2004). 
With the relative velocities and positions at the beginning and end of a time step, 
the closest distance of a pair during the step is calculated using a third order interpolation 
(Eq.~(11) of Chambers, 1999). 
In order to judge whether a particle is in a close encounter or not,
we first define the critical distance for each particle.
The critical distance is defined as the larger
of (1) $n_{1} R_{\rm H}$ or (2) $n_{2}v_{\rm ep} \Delta t$,
where $n_{1}$ and $n_2$ are numerical factor 
(we use $n_{1} = 4.5$ and $n_2 = 1.0$), $R_{\rm H}$ is the Hill radius,
and $v_{\rm ep} = v_{\rm kep}\sqrt{e^2+i^2}$ is the epicyclic velocity of the particle
with $v_{\rm kep}$, $e$, and $i$ being the Kepler velocity, eccentricity,
and inclination, respectively.
If the closest distance of a pair is smaller than the sum of the critical distances,
we regard this pair to be in a close encounter.  
In the original SyMBA code (Duncan et al., 1998), only 
the first condition is used.  However, the second condition is necessary 
if the relative velocity is much larger than the mutual escape velocity;  such encounters usually occur 
near resonances of the giant planets or between small planetesimals stirred by large protoplanets. 
In the Mercury code (Chambers, 1999), the second condition uses the largest $v_{\rm kep}$ 
in all particles instead of each body's $v_{\rm ep}$ which we apply. 
Although this criterion of the Mercury code gives better energy conservation (for a same $n_2$),  
it also results in too many pairs being identified as involved in close encounters during the runaway growth stage, 
where the number density of particles is large.
If the number density is not large, as is the case in the late stage, use of the Mercury's criterion is recommended
(although we do not use it even in the late stage in this paper). 

All the above computations are performed in parallel (if  $N > 1000$), 
while particles in close encounters are collected on a single cpu during the domain decomposition; the procedure 
which distributes the workload, and hence the remaining particles, amongst processors.
Then, multistep time stepping is done until all particles proceed to the end 
of the original time step, following the procedure described in Duncan et al. (1998).  
The closest mutual distance in a substep is always estimated by interpolation using the information of both the 
beginning and the end of the step. 
This guarantees time symmetry of the integrator which is essential to avoid secular errors. 
The computational time for the substep procedure is small enough as it is 
simply proportional to the number of particles in close encounters.
We set  the ratio of the cut-off radius to that of the next order to be 2.08 and the number of next order substeps 
in the current substep to be 3 after Duncan et al. (1998).

We assume that two bodies merge when they partially overlap. 
We assume the density of particles to be 2 g cm$^{-3}$.  
Since we do not apply the artificial enhancement of radii (Kokubo and Ida, 2002, Morishima et al., 2008), 
the accretion time scale we will show is realistic. 
A particle is assumed to collide with the Sun when either the heliocentric distance is less than 0.1 AU
or the semimajor axis is less than 0.25 AU. A particle is assumed to escape from the system 
when the heliocentric distance is larger than 100 AU.
The energy error after completion of accretion is typically $\Delta E/E = 10^{-5}$--$10^{-3}$. 
Note that energy changes due to merging or gas forces can be explicitly calculated and are subtracted in the above error.
The largest portion of the error usually accumulates after the runaway growth stage where a large number of excited 
small particles experience mutual encounters.
Therefore, the energy error primarily depends on how long small particles survive. 

\subsection{Effects of gas}
In this section, we review the forces due to the nebular gas.  
The acceleration due to the gas is incorporated in the integrator by placing the 
half kicks before and after the Kepler drift operation.
This symmetric placement once again avoids secular errors coming from the inclusion of gas forces.

\subsubsection{Aerodynamic drag}
The aerodynamic drag force per unit mass (or acceleration) is given by (Adachi et al., 1976)
\begin{equation}
\mbox{\boldmath $f$}_{\rm drag} = 
-\frac{1}{2m}c_{\rm D}\pi r_{\rm p}^2 \rho_{\rm gas} 
|\mbox{\boldmath $v$}_{\rm rel}|\mbox{\boldmath $v$}_{\rm rel},
\end{equation}
where $m$ and $r_{\rm p}$ are the mass and the radius of a particle,
$c_{\rm D}$ is the numerical coefficient (assumed to be 2 in the present paper), 
 $\rho_{\rm gas}$ is the gas density, and 
$\mbox{\boldmath $v$}_{\rm rel} = \mbox{\boldmath $v$} 
- \mbox{\boldmath $v$}_{\rm gas}$
is the velocity of a particle relative to the gas velocity.
In the cylindrical coordinate $(r, \theta, z)$,
the gas velocity in the $\theta$ component is obtained by the force balance in the $r$ component as 
\begin{equation}
\frac{v_{\rm gas}^2 (r,z)}{r} = 
\frac{c^2}{r} \left(\frac{\partial \ln \rho_{\rm gas}}{\partial \ln r} + \frac{\partial\ln T}{\partial\ln r}\right)
+ \frac{GM_{\odot}}{(r^2+z^2)^{3/2}}r
-  f_{{\rm glob}, r} (r,z), \label{eq:vgas}
\end{equation}
where $c$ is the isothermal sound velocity, $T$ is the gas temperature, $G$ is the gravitational constant,
and $M_{\odot}$ is the mass of the central star (assumed to be the solar mass).
The left hand side of Eq.~(\ref{eq:vgas}) represents the centrifugal force
while the first, second, and third terms of the right hand side 
represent the pressure gradient,  the gravity of the central star,
and the self-gravity of the gas disk (Eq.~(\ref{eq:fglr})), respectively.
Assuming the gas is isothermal in the $z$ direction and neglecting the self gravity 
of the gas, the gas density is approximately given by
\begin{equation}
\rho_{\rm gas}(r, z) = \left(\frac{\Sigma_{\rm gas}}{\sqrt{2\pi}h}\right)
\exp{\left(-\frac{z^2}{2h^2}\right)},
\end{equation}
where $\Sigma_{\rm gas}$ is the surface density of the gas and
$h = c/\Omega_{\rm kep}$ is the scale height, with the Keplerian frequency 
$\Omega_{\rm kep}$. 
If we assume $\Sigma_{\rm gas} \propto r^{-p}$, $T \propto r^{-q}$,
we obtain, 
\begin{equation}
\frac{\partial \ln \rho_{\rm gas}}{\partial \ln r} 
= - p + \left(\frac{q}{2} - \frac{3}{2}\right)
\left(1-\left(\frac{z}{h}\right)^2\right).
\end{equation}
We adopt the temperature profile of an optically thin disk, $T =$ 280 $(r/{\rm 1\hspace{0.3em}AU})^{-1/2}$K ($q = 1/2$), 
after Hayashi et al. (1981).  
We will explain how $\Sigma_{\rm gas}$ and $p$ are given in Sec.~2.3.

The growth rate of a protoplanet is proportional to $e^{-2}$, 
where $e$ is the orbital eccentricity of planetesimals, and the balance between 
viscous stirring by protoplanets and damping due to gas drag gives 
$e \propto m^{1/18}$ (Eq.~(25) of Kokubo and Ida, 2000).
The initial planetesimal mass $m_0$ in our simulations is $\sim 10^{25}$ g.
On the other hand, the size of planetesimals $m_{\rm GI}$ formed by gravitational instability 
is estimated to be several order of magnitude smaller (Goldreich and Ward, 1973).
Therefore, the growth time scale of protoplanets will be significantly overestimated in our model.
In order to reproduce the growth rate of protoplanets 
surrounded by small planetesimals ($\sim m_{\rm GI}$), we artificially enhance the gas drag force on planetesimals 
by a factor of $(m/m')^{2/3}$ in $N$-body simulations
if the particle mass $m$ is smaller than $m_{\rm 1}$ ($> m_{0}$).
Here the mass $m'$ is given as
\begin{equation}
\frac{\log{m'}-\log{m_{\rm GI}}}{\log{m_1}-\log{m_{\rm GI}}}
= \frac{\log{m}-\log{m_0}}{\log{m_1}-\log{m_0}}. 
\hspace{1em}({\rm for } \hspace{1em} m < m_1)
\end{equation}
With this scheme, a bunch of small planetesimals of mass $m'$ 
are represented by a single super planetesimal of mass $m$ (e.g., McNeil et al., 2005).
The mass $m_1$ should be as small as possible in order to avoid artificial clean up of 
large planetesimals by protoplanets.
We use $m_{\rm 1} = 0.01$ $M_{\oplus}$ as a compromise, and $m_{\rm GI} = 10^{19}$ g.
The migration and damping time scales with the artificial enhancement of the gas drag are shown in Fig.~1 
for the case of $e = 0.05$ and $m_0 = 0.0025$ $M_{\oplus}$. 
\marginpar{\textbf{[Fig.~1]}}

\subsubsection{Tidal interaction between a protoplanet and a nebular disk}
We use the formulation of Tanaka et al. (2002) for Type I migration rate. 
For the damping rates of orbital eccentricity $e$ and inclination $i$, 
we use the formulation given in Tanaka and Ward (2004). 
We make corrections in these formulations for the case of large $e$ and $i$
after Papaloizou and Larwood (2000) and Cresswell et al. (2007).

Tanaka et al. (2002) derived the torque due to tidal interaction between 
a protoplanet and a three-dimensional baroclinic disk, for the case of $e=i=0$. 
The decay time scale of the semimajor axis $a$ is given as 
\begin{equation}
\tau_{\rm tid1} = -\frac{a}{\dot{a}} 
= -\frac{1}{2}\frac{v_{\rm kep}}{f_{{\rm tid1},\theta}}
= (2.7+1.1p)^{-1} 
\frac{M_{\odot}}{m}
\frac{M_{\odot}}{\Sigma_{\rm gas} a^2}
\left(\frac{c}{v_{\rm kep}} \right)^2 \Omega_{\rm kep}^{-1}, \label{eq:type1}
\end{equation}
where $f_{{\rm tid1}, \theta}$ is the acceleration in the azimuthal direction. 

Tanaka and Ward (2004) extended their work to the case of nonzero $e$ and $i$ ($\ll h/a$)
and obtained the rate for damping of $e$ and $i$ due to the tidal interaction. 
The $r$, $\theta$, and $z$ components of the damping force per unit mass 
are given as (a typo was corrected; see Ogihara et al., 2007)
\begin{eqnarray}
f_{{\rm tid},r} &=& \frac{1}{\tau_{\rm wave}} 
\left[0.104[v_{\theta} - v_{\rm kep}'(r)] + 0.176 v_r \right], 
\label{eq:ftidr}  \\ 
f_{{\rm tid},\theta} &=&  \frac{1}{\tau_{\rm wave}} 
\left[-1.736[v_{\theta} - v_{\rm kep}'(r)] + 0.325 v_r \right], 
\label{eq:ftidth} \\ 
f_{{\rm tid},z} &=&
\frac{1}{\tau_{\rm wave}} 
\left[-1.088v_z + 0.871z \Omega_{\rm kep}(r) \right], \label{eq:ftidz}
\end{eqnarray}
where characteristic time of the orbital evolution, 
$\tau_{\rm wave}$, is given by  
\begin{equation}
\tau_{\rm wave} = 
\left(\frac{m}{M_{\odot}} \right)^{-1}
\left(\frac{\Sigma_{\rm gas} a^2}{M_{\odot}}\right)^{-1}
\left(\frac{c}{v_{\rm kep}} \right)^4 \Omega_{\rm kep}^{-1}. \label{eq:twave}
\end{equation}
In Eqs.~(\ref{eq:ftidr})--(\ref{eq:ftidz}), the velocity components of a particle are given as 
$\mbox{\boldmath $v$} = (v_r, v_{\theta}, v_z)$ and 
 $v_{\rm kep}'(r) = (v_{\rm kep}^2 - rf_{{\rm glob},r}(r,0) )^{1/2} $ is the Keplerian velocity of solid bodies 
including the gravity of the gas disk. 
Note that all the components of the damping force become zero when $e=i=0$.

The formulation of Tanaka and Ward (2004) is limited to small $e$ and $i$ ($\ll h/a$), 
or that the planet motion relative to the gas remains subsonic.
From a two dimensional linear theory, Papaloizou and Larwood (2000) 
found that the time scales $\tau_{\rm tid1}$ and $\tau_{\rm wave}$ (Eqs.~(\ref{eq:type1}) and (\ref{eq:twave})) 
increase with increasing $e$ by factors of $g_{\rm tid1}$ and  $g_{\rm wave}$, respectively: 
\begin{eqnarray}
g_{\rm tid1} =  
\frac{1+ (x/1.3)^5}{1 - (x/1.1)^4}, \label{eq:fmig} \\
g_{\rm wave} = 1 + \frac{1}{4}x^3. \label{eq:fwave}
\end{eqnarray}
Here the original $x$ used in Papaloizou and Larwood (2000) is  $x_{PL} = ae/h$.
On the other hand, we use a modified value as  
\begin{equation}
x = \frac{a\sqrt{(e^2+i^2)}}{2h},  \label{eq:modx}
\end{equation}
in order to reproduce similar results of simulations of Cresswell et al. (2007) (see the next paragraph).
The evolution of the orbital eccentricity with the corrections for large $e$ is given as 
\begin{equation}
\frac{1}{e} \left( \frac{de}{dt} \right)_{\rm tid}  
= - \frac{0.78}{\tau_{\rm wave}g_{\rm wave}} 
\label{eq:dedt}.
\end{equation}
This equation corresponds to Eq.~(45) of Tanaka and Ward (2004), 
but the correction factor is multiplied by $\tau_{\rm wave}$. 
From Eqs.~(\ref{eq:type1}), (\ref{eq:fmig}), and (\ref{eq:dedt}), the evolution of the semimajor axis 
due to the tidal interaction is given by, 
\begin{equation}
\frac{1}{a} \left(\frac{da}{dt}\right)_{\rm tid}   
= -\frac{1}{\tau_{\rm tid1}g_{\rm tid1}} 
- \frac{1}{e} \left( \frac{de}{dt}\right)_{\rm tid}  
\left(\frac{2e^2}{\sqrt{1-e^2}}\right),
\label{eq:dadt}
\end{equation}
where the first term in the right hand side comes from exchange of orbital 
angular momenta while the second term represents the migration rate 
due to energy dissipation. 

Cresswell et al. (2007) conducted hydrodynamic simulations for migration of a
20 Earth mass planet with nonzero $e$ and $i$ embedded in a gas disk.
They found good agreement with Tanaka et al. (2002) and Tanaka and Ward (2004) for small $e$ and $i$ ($< 0.1$),
while the damping time scale $e/\dot{e}$ is proportional to $e^3$ for large $e$ 
as predicted by Papaloizou and Larwood (2000) (Eq.~(\ref{eq:fwave})).
They also showed the same dependence of the inclination decay rate on $i$, 
so we first simply replace $e$ by $(e^2 + i^2)^{1/2}$ in Eq.~(\ref{eq:modx}). 
With the original $x_{\rm PL}$, Eq.~(\ref{eq:dadt}) gives outward migration for $ae/h > 1.1$. 
Cresswell et al. (2007) also found that the sign of the torque changes to positive ($g_{\rm tid1} < 0$) when  
$ae/h > 1.6$, but did not find any indication of outward migration. This means that the absolute value of the second term 
in the right hand side of Eq.~(\ref{eq:dadt}) 
is always larger than the first term for large $e$. It is unclear whether outward migration really occurs or not
and we need to wait for future studies. In the present paper, we use the modified $x$ (Eq.~(\ref{eq:modx}))
to reproduce similar dependences of $\dot{e}$ and $\dot{a}$ on $e$ and $i$ to those obtained in Cresswell et al. (2007).
The migration and damping time scales ($e/\dot{e}$ and $a/\dot{a}$) with various $e$ are shown in Fig.~1. 
Although there are some differences between 
the time scales shown in Fig.~1 and those from Cresswell et al. (2007), these do not result in significant 
differences in our $N$-body simulations unless the time scales become too large or negative.
Practically, large protoplanets obtain large $e$ when they are trapped in secular resonances. 
If outward migration really occurs, protoplanets can be removed from secular resonances which migrate inward.

\subsubsection{Global nebular force and secular resonances}
We calculate the gravity of the gas disk after Nagasawa et al. (2000).
The $r$ and $z$ components of the nebular gravitational force are given as
\begin{eqnarray}
f_{{\rm glob}, r}(r,z) &=& -2G \int^{\infty}_{-\infty} \int_{r'}\frac{r'}{r}
\frac{\rho_{\rm gas}(r',z')}{\sqrt{(r+r')^2 + (z-z')^2}} \label{eq:fglr}
\nonumber \\
&& \times \left[\frac{r^2-r'^2 -(z-z')^2}{(r-r')^2 -(z-z')^2}E(k) + K(k)\right]
dr'dz'
\nonumber \\
f_{{\rm glob}, z}(r,z) &=& -4G  \int^{\infty}_{-\infty} \int_{r'}
\frac{\rho_{\rm gas}(r',z')r'(z-z')}{\sqrt{(r+r')^2 + (z-z')^2}}
\left[\frac{E(k)}{(r-r')^2 +(z-z')^2}\right]dr'dz',
\end{eqnarray}
where $K(k)$ and $E(k)$ are elliptic integrals of the first and second kind, 
and $k$ is given by
\begin{equation}
k = \sqrt{\frac{4rr'}{(r+r')^2 + (z-z')^2}}.
\end{equation}
These components of the force are initially tabulated in a ($r,z$) grid, and interpolated in simulations. 
A secular resonance occurs when precession rates of two orbiting bodies coincide
(Heppenheimer, 1980; Ward, 1981; Nagasawa et al., 2000). 
We analytically calculate precession rates of planetesimals and the gas giant planets for $e$, $i$ $\ll 1$,
 following Appendix~A of Nagasawa et al. (2000).   
Then, we obtain the locations of the secular resonances with Jupiter ($\nu_5$) and Saturn ($\nu_6$)  (e.g., Fig.~2). 

\subsection{Input parameters}
We adopt four different initial spatial distributions of planetesimals,
four different dissipation time scales of the nebular gas, and four different types of orbits of Jovian planets.
In total, therefore, we made 64 runs with different combinations of parameters. Each run takes one to three cpu months 
until completion of accretion.

\subsubsection{Initial spatial distribution of planetesimals}
We initially place 2000 equal-mass planetesimals between 0.5 and 4.0 AU.
The initial surface density distribution of planetesimals is given by
\begin{equation}
\Sigma_{\rm solid} = \Sigma_{{\rm solid},0}\left(\frac{r}{\rm 1\hspace{0.3em} AU}\right)^{-p}, 
\hspace{0.5em} ({\rm for \hspace{0.5em} 0.5 \hspace{0.3em} AU} \le r \le 4.0\hspace{0.3em}{\rm AU})
\end{equation}
where $\Sigma_{\rm solid,0}$ is $\Sigma_{\rm solid}$ at 1 AU. We adopt $p = 1$ and 2 in simulations.
The initial total solid mass is given as 
\begin{equation}
M_{\rm T} = \int_{\rm 0.5 \hspace{0.2em} AU}^{\rm 4.0 \hspace{0.2em} AU} 2 \pi r \Sigma_{\rm solid}dr.
\end{equation}
We adopt $M_{\rm T} = 5$ and $10$ $M_{\oplus}$. 
In the case of $M_{\rm T} = 5$ $M_{\oplus}$, 
$\Sigma_{{\rm solid},0} =$ 6.1 and 10.2 gcm$^{-2}$ for $p = 1$ and 2, 
respectively.  
For comparison, the minimum mass solar nebula (MMSN)
by Hayashi et al. (1981) has $\Sigma_{\rm solid,0} = 7.1$ gcm$^{-2}$,  $p = 1.5$, and $M_{\rm T} = 4.3$ $M_{\oplus}$.

\subsubsection{Dissipation  time scale of nebular gas}
In the present paper, we limit ourselves to the case where 
the surface density of gas dissipates exponentially in time and uniformly in space:
\begin{equation}
\Sigma_{\rm gas}(r, t) = 
\Sigma_{\rm gas, 0}\left(\frac{r}{\rm 1 \hspace{0.3em} AU}\right)^{-p} \exp{\left(-\frac{t}{\tau_{\rm decay}}\right)}, 
\hspace{0.5em} ({\rm for \hspace{0.5em} 0.1 \hspace{0.3em} AU} \le r \le 36.0 \hspace{0.3em} {\rm AU})
\end{equation}
where $\Sigma_{\rm gas,0}$ is $\Sigma_{\rm gas}$ at 1AU and $t=0$ and $\tau_{\rm decay}$ is the time scale for gas decay.
We adopt four different decay times: $\tau_{\rm decay} = 1,2,3$, and 5 Myr.
We assume that $\Sigma_{{\rm gas},0} = 2000$ and 3366 gcm$^{-2}$ for $p$
 = 1 and 2, respectively.   
The gas-to-solid ratio is constant between 0.5 AU and 4.0 AU in the beginning of simulations and is 329 and 165
for $M_{\rm T} = 5$ and 10 $M_{\oplus}$, respectively.
For comparison, MMSN has $\Sigma_{\rm gas, 0} = 1700$ gcm$^{-2}$ (Hayashi et al., 1981).
We do not consider a gap around Jupiter's orbit in the present paper. 
The formation of a gap does not significantly alter the time of the sweeping 
secular resonance's passage in the gaseous region (Nagasawa et al., 2005). 
However, orbital evolutions of asteroidal bodies in a gas-free gap will be different, in particular, if a gap is very wide.

\subsubsection{Orbits of Jovian planets}
We introduce Jupiter and Saturn after $\tau_{\rm decay}$ from the beginning of a simulation, assuming that 
they form via core accretion (Bodenheimer and Polack, 1986;  Ikoma et al., 2000; Lissauer et al., 2009).  
Rapid formation of Jovian planets via disk instability is unlikely for our solar system because efficient cooling 
for clump formation is only possible at $r > 100$ AU (Boley, 2009). 

We define $a_{\rm J}$ and $e_{\rm J}$ to be the semimajor axis and the orbital eccentricity of Jupiter, 
and $a_{\rm S}$ and $e_{\rm S}$ to be the corresponding orbital elements for Saturn. 
We adopt four types of orbits for Jupiter and Saturn at their formation 
and use the same notation introduced in O'Brien et al. (2006) and Raymond et al. (2009), 
although we have slightly different orbits:
\begin{itemize}
 \item EJS ("Eccentric Jupiter and Saturn"). Jupiter and Saturn are placed on their current orbits:
 $a_{\rm J} = 5.20$ AU, $e_{\rm J} = 0.048$, $a_{\rm S}$ = 9.55 AU, and $e_{\rm S} = 0.056$.
 \item CJS ("Circular Jupiter and Saturn"). These are the initial conditions used in the Nice model (Tsiganis et al., 2005).
 Jupiter and Saturn are radially more confined with circular orbits: 
 $a_{\rm J} = 5.45$ AU, $e_{\rm J} = 0.0$, $a_{\rm S}$ = 8.18 AU, and $e_{\rm S} = 0.0$. The mutual inclination is 
 0.5 degrees.
 \item EEJS ("Extra Eccentric Jupiter and Saturn"). The same as EJS except a higher eccentricity of Jupiter: $e_{\rm J} = 0.1$.  
 \item CJSECC. Jupiter and Saturn have the CJS semimajor axes with higher eccentricities: $e_{\rm J} = 0.05$  and $e_{\rm S} = 0.05$. 
\end{itemize}

The EJS and EEJS simulations assume that Jupiter and Saturn formed at their current locations.
On the other hand, the Nice model suggests that Jupiter migrated inward and Saturn migrated outward from their 
original CJS orbits via interactions with Kuiper Belt objects leading to a sudden change to the EJS orbits when 
Jupiter and Saturn crossed their 1:2 orbital resonance (Tsiganis et al., 2005). 
Asteroidal and cometary bodies strongly perturbed by this event are considered to have caused 
the Late Heavy Bombardment at $\sim$ 3.85 billion years ago; well after formation of the terrestrial planets 
(Gomes et al., 2005; Strom et al., 2005).
Initially higher $e_{\rm J}$ and $e_{\rm S}$ in the EEJS and CJSECC simulations
than those in the EJS and CJS simulations, respectively, are considered because $e_{\rm J}$ and $e_{\rm S}$ decrease via interaction with planetesimals (Sec.~3).

We only take into account the gravity of other bodies and the global nebular force on motion of Jupiter and Saturn.
We neglect any further interaction between a gas disk and the giant planets. Details of orbital evolution of a 
giant planet, which opens a gap in a gas disk, have a complicated dependence on disk properties, 
such as shape of a gap and disk mass, for which explicit and reliable formulae are not yet available.
Nevertheless, some basic trends, which should be included in future studies, started to be clarified as follows.
Analytic (Goldreich and Sari, 2003; Ogilvie and Lubow, 2003) and numerical (D'Angelo et al., 2006) studies 
indicate that eccentricity growth for a Jupiter-mass planet occurs together with the saturation of the 
corotation resonance in a low-viscosity disk. 
The migration time scale of a Jovian planet is much longer than the viscous diffusion time scale and even outward 
migration occurs, either when the gap is clean and the orbit is eccentric (D'Angelo et al., 2006) or when the gap 
is not deep enough (Crida and Morbidelli, 2007). 
A long migration time scale of Jupiter due to disk-sculpturing by Saturn is also pointed out by Morbidelli and Crida (2007). 

\section{Results}
\subsection{Time evolution}
\subsubsection{Case with EJS orbits}
Figures 2-4 show an example of the time evolution for the case with the present orbits of Jupiter and Saturn:
snapshots on the plane of the semimajor axis versus the orbital eccentricity (Fig.~2), evolution 
of the total masses and the mean eccentricities of small and big particles (Fig.~3), and evolution of 
the masses and orbital excursions of surviving planets (Fig.~4).
\marginpar{\textbf{[Figs.~2-4]}} 
The gas decay time is $t_{\rm decay} = 1$ Myr, the initial total mass is 5 $M_{\oplus}$, the power-law index 
for the surface density is $p=2$.  
In the present paper, we distinguish between a big particle, or (proto)planet, defined as a body with 
$m > 2\times 10^{26}$ g ($\sim 0.03$ $M_{\oplus}$), and smaller bodies or planetesimals. 
This border mass is slightly smaller than Mercury's mass ($3.3 \times 10^{26}$ g), 
and one order of magnitude smaller than the isolation mass at 1AU (Kokubo and Ida, 2002). 

We follow growth of protoplanets without Jupiter and Saturn, until $t_{\rm decay}$ from the beginning of the simulation. 
We first observe the mass loss ($\sim 0.2$ $M_{\oplus}$ at $10^5$ yr; see Fig.~3)
by the migration of the smallest planetesimals due to gas drag at the inner most region.
Protoplanets grow rapidly in the inner regions while many smaller bodies remain unaccreted in the outer regions. 
Protoplanets first tend to collide with smaller planetesimals because their orbital eccentricities are smaller.
At 1 Myr, the smallest particles significantly deplete at $r < 2$ AU, while some larger
planetesimals still survive there (Fig.~2). 
Once a planet reaches a certain size, it starts rapid inward migration (Type I migration).  
Since the growth time is shorter for smaller $r$, significant mass depletion first 
occurs in the inner disk. With increasing time, the mass depletion propagates outward 
(see Ida and Lin (2008) for more details).
The mass loss due to Type I migration is easily recognized in the upper panel of Fig.~3,
as we see stepwise depletion in the total mass. More than half of the total mass is lost by this mechanism until 3 Myr. 
The half of the total mass corresponds to the initial total mass inside of $\sim$1.5 AU.
The three planets, which survive till the end, are only lunar-size at 1 Myr and located outside of 1.2 AU (Fig.~4). 
There are other larger bodies located inside of 1.2 AU at 1 Myr, but they will not survive Type I migration.

At 1 Myr, we introduce Jupiter and Saturn with their present orbital elements.
Their formation causes a sudden increase of orbital eccentricities of nearby particles, particularly those in mean and 
secular resonances, while such a jump does not appear for inner planets (see the lower panel of Fig.~3).  
The locations of the $\nu_5$ and $\nu_6$ resonances at $t = 1$ Myr are 3.10 and 3.34 AU, respectively. 
These resonances, in particular $\nu_5$, effectively enhance orbital eccentricities of planetesimals and protoplanets, 
and these bodies migrate inward as the gas drag and the tidal interaction damp their eccentricities.  
At 3-5 Myr, the mean eccentricities of big and small bodies are comparable (Fig.~3) as many protoplanets are 
also involved in the $\nu_5$ resonance.  Near the resonance, the surface density of solid material is highly enhanced
and relative velocities of solid bodies are small due to orbital alignments (Nagasawa et al., 2005).  
Consequently, protoplanets grow very quickly (Fig.~4).  Collisions between protoplanets also commonly occur at this stage.

When the gas density becomes so low that inward migration of planetesimals and protoplanets due to the gas drag or 
the tidal interaction is slower than the migration rate of the secular resonance, these bodies are no longer swept by 
the resonance (Nagasawa et al., 2005).  After the passage of the $\nu_5$ resonance, 
orbital eccentricities of planets are reduced first by tidal interaction with a remnant 
gas disk. The damping due to gas is effective as long as the damping time 
 is shorter than the gas decay time ($\tau_{\rm damp} < \tau_{\rm decay}$). 
For an Earth-size planet at 1 AU, this condition is satisfied until $\sim$ 7--8 $\tau_{\rm decay}$. 
Indeed, we see a rapid decrease of the mean eccentricity of big bodies at 5-7 Myr in Fig.~3. 
Later, the damping due to dynamical friction of planetesimals further reduces the eccentricities of planets. 	 
Its effect is determined by the total mass and the mass distribution of planetesimals (Morishima et al., 2008).
Small bodies contain $\sim 20\%$ of the total mass at 10 Myr and this amount is likely to be sufficient 
to achieve energy equipartition between big and small bodies as long as the planet-planet interaction is negligible. 
The typical size of remnant planetesimals is the lunar size or slightly smaller. 
There are only a few remnant planetesimals with the initial size because the smallest planetesimals had small orbital eccentricities due to 
strong gas drag and efficiently merged with larger bodies. The equilibrium eccentricity of Earth-size bodies 
due to dynamical friction of lunar-size planetesimals is estimated to be $\sim$ 0.03 (Morishima et al., 2008),
and we obtain a quite consistent value after most of the remnant planetesimals merge with planets. 

Through the above processes, three planets form near 1 AU at the end of this simulation.
The largest planet has 1.4 $M_{\oplus}$ while both of other two have 0.3 $M_{\oplus}$. 
The locations of these three planets resemble those of the current terrestrial planets
(except we do not have a Mercury analog), although the masses are somewhat different. 
Orbital eccentricities are as small as those for the current planets. The most serious issue for this simulation is that 
the last of the potentially Moon-forming impacts occurs too early (at 13.5 Myr; see Sec.~3.2 for the definition) 
when compared with that suggested from isotope records (Touboul et al., 2007).  

\subsubsection{Case with CJS orbits}
Figures~5-7 show the time evolution for the case with circular orbits of Jupiter and Saturn, which are given the initial 
configuration of the Nice model. \marginpar{\textbf{[Figs.~5-7]}} 
Except for the orbits of the gas giant planets, all other parameters are the same as those used in the EJS simulation 
in Sec.~3.1.1 (Figs.~2-4).
If the orbits of the gas giant planets are nearly circular, the effects of the secular resonances are much weaker 
than the case of the EJS orbits. Consequently, a large fraction of particles stay in the asteroidal region for a long 
period of time.
Nevertheless, the orbital eccentricities of the gas giant planets can not be completely zero when there are two (or more)  
gas giant planets. Therefore, resonances gradually sculpture orbits of particles in the asteroidal region. 
Some enhancements at the 3:1 mean motion resonance (2.6 AU) and the $\nu_6$ secular resonance (3.2 AU) are seen at 
3 Myr (Fig.~5). These enhancements gradually propagate to non-resonant locations via mutual gravitational 
interactions (Raymond et al., 2006). 
The mass transfer from the asteroidal region to the terrestrial region slowly occurs as excited asteroidal bodies encounter 
and collide with inner planets.

At 3 Myr, there are about 15 Mars-size protoplanets in the terrestrial region.  
The fraction of small particles is very small in the terrestrial region whereas 
small particles are still dominant in mass in the asteroidal region.
This situation is similar to some of the initial conditions adopted in previous $N$-body simulations (e.g., Raymond et al., 2006).
In our simulation, however, the mean orbital separation between protoplanets is $\sim$ 20 Hill radius at 3 Myr 
while previous late stage simulations usually assume the orbital separation to be $\sim 10$ Hill radius, based on the oligarchic growth model of Kokubo and Ida (1998).  The larger separation in our simulation
is due to Type I migration, which makes inner protoplanets start to migrate inward 
earlier than outer planets  (McNeil et al., 2005). With the large orbital separation and the orbital stabilization due to 
the nebular gas, mutual collisions between protoplanets are rare until $\sim$ 10 Myr.  

After 10 Myr, mutual collisions between protoplanets start
to occur, and the orbital separation increases as the number of protoplanets decreases.
Strictly speaking, the orbital separation normalized by the mutual Hill radius, and averaged over pairs of planets,
increases nearly logarithmically with time (from $\sim 20$ at 3 Myr to $\sim 60$ at 800 Myr), 
and this is probably because the stable time scale of a planetary system exponentially increases with 
the normalized separation (Chambers et al., 1996; Ito and Tanikawa, 1999; Yoshinaga et al., 1999). 
Through giant impacts, an Earth-size planet forms slightly inside of 2 AU. 
This planet highly enhances orbital eccentricities of asteroidal bodies and their number gradually decreases by either 
hitting this planet or escaping from the system after encounters with Jupiter.

There are still a large number of remnant planetesimals at 100 Myr, but not at 300 Myr (Figs.~5 and 6).
Therefore, damping due to planetesimals is expected until giant impacts complete at around 100 Myr.
In this simulation, the last potentially Moon-forming impact occurs at 139.5 Myr on the outermost planet (Fig.~7).
Nevertheless, the mean orbital eccentricity in the end of the simulation ($\sim$ 800 Myr) is not small,  
as the two middle planets are still strongly interacting (Fig.~7). 
These two planets are stirred by the innermost and outermost planets which have larger masses than the middle planets. 
This simulation succeeded in reproducing the very late giant impact, but failed to reproduce
the nearly circular orbits and the radial mass concentration of the current system.

\subsection{Comparison with constraints}
In this section, we examine the properties of planetary systems formed in all runs and compare them with those for 
our own system. We consider the following three properties of systems (see also Raymond et al., 2009).
\begin{enumerate}
\item The spatial mass distribution.
The radial mass concentration statistic, $S_c$, is defined as (Chambers, 2001)
\begin{equation}
S_c = {\rm max}\left(\frac{\sum m_j}{\sum m_j \log_{10}(a/a_j)}\right),
\end{equation}
where $m_j$ and $a_j$ are the mass and the semimajor axis of each planet, 
and the subscript $j$ is given in the order of $a_j$. 
The function in the parenthesis is calculated for $a$ throughout the terrestrial planet region, with $S_c$ being 
the maximum value. This quantity increases with radial mass concentration. 
We use two additional parameters as constraints. The first one is the minimum orbital separation between 
neighboring planets $b_{\rm min}$ normalized by their mutual Hill radius 
$r_{{\rm H},j} = 0.5(a_j+a_{j+1})[(m_j+m_{j+1})/(3 M_{\odot})]^{1/3}$: 
\begin{equation}
b_{\rm min} = {\rm min}\left(\frac{a_{j+1}-a_j}{r_{{\rm H},j}}\right).
\end{equation}
The second being the mass weighted mean semimajor axis of planets 
\begin{equation}
a_m = \frac{\sum a_j m_j}{M_{\rm T,final}},
\end{equation} 
where $M_{\rm T,final} = \sum m_j$ is the final total mass of planets. 
For the terrestrial planets,  $S_c = 89.9$, $b_{\rm min} = 26.3$ 
(the normalized orbital separation between the Earth and Venus), and $a_m =$0.90 AU. 
 
\item The deviation from circular and coplanar orbits.
The orbital excitation of terrestrial planets is characterized by the angular momentum deficit, $S_d$ (Laskar, 1997):
\begin{equation}
S_d = \left(\frac{\sum m_j \sqrt{a_j} \left( 1-\cos(i_j)\sqrt{1-e_j^2} \right)}{\sum m_j \sqrt{a_j}}\right),
\end{equation}
where $e_j$ and $i_j$ are the orbital eccentricity and inclination of each planet with respect to the 
fiducial plane of an initial planetesimal disk. We also measure the mass weighted mean eccentricity:
\begin{equation}
e_m = \frac{\sum e_j m_j}{M_{\rm T,final}}.
\end{equation}
We use the orbital elements averaged over a few Myr.  
For the terrestrial planets,  $S_d = 0.0018$ and $e_m = 0.038$ (Quinn et al., 1991). 

\item The timing of the Moon-forming impact. Based on simulations for giant impacts 
and accretion of moons (Cameron and Benz, 1991; Ida et al., 1997; Canup et al., 2001; Canup, 2004; 2008), 
we roughly regard an impact as potentially Moon-forming, if  
a) the total mass of the impactor and the target is $> 0.5$ $M_{\oplus}$, b) the impactor's mass 
($<$ the target's mass) is $> 0.05$ $M_{\oplus}$, and 
c) the impact angular momentum is $> 0.5$ $L_{\rm EM}$, where $L_{\rm EM}$ is the current total angular momentum of 
the Earth-Moon system (i.e., the Earth's spin angular momentum and the Moon's orbital angular momentum). 
We define the time of the last potentially Moon-forming impact in a run to be $t_{\rm imp}$. 
The time of the actual Moon-forming impact is estimated as $t_{\rm imp} = 50$-150 Myr (Touboul et al., 2007;  
All\'{e}gre et al., 2008). 
\footnote[2]{Smaller  $t_{\rm imp} (\simeq 30$ Myr) is not ruled out from the terrestrial W/Hf records because 
the degree of metal-silicate equilibration at the Moon-forming impact is still in debate (Jacobsen et al., 2009).
In this case, however, one needs an alternative interpretation for the prolonged differentiation of the Moon 
(Touboul et al., 2007), other than the late Moon-forming giant impact.}
The small content of iron in the Moon is explained when the impact velocity is as slow as the mutual escape 
velocity (Canup, 2004, 2008); this constraint will be also discussed (Figs.~13 and 16). 
\end{enumerate}

Raymond et al. (2009) considered two additional constraints.
The first one is the absence of planetary bodies ($> 0.05$ $M_{\oplus}$) in the asteroidal region ($>$ 2 AU).
They showed that bodies which are too massive destroy the observed radial variation of the asteroidal taxonomic types 
(Gradie and Tedesco, 1982).
If massive bodies exist in the asteroidal region, this system usually has too large $a_m$ or too small $S_c$, 
because we do not exclude a massive body in the asteroidal region to be defined as a planet. 
Therefore, this constraint is automatically included in our first constraint. 
Their second additional constraint is the Earth's water content. 
They showed that the mass supply from asteroidal bodies to terrestrial planets decreases 
with increasing eccentricity of Jupiter. As a result, the water content of an Earth-analog is insufficient  
in their EJS and EEJS simulations. 
On the other hand, in our EJS and EEJS simulations with the nebular gas, a large fraction of mass of 
an Earth-analog comes from the asteroidal region, while protoplanets which formed from planetesimals originally
in the terrestrial region spiral into the Sun by Type I migration (Sec.~3.1.1). 
Probably, the problem here is rather that we need to consider how to remove excess water from our Earth-analogs.

\subsubsection{Comparison between EJS and CJS simulations}
The properties of the planetary systems from the EJS and CJS simulations are summarized in Table~1.
\marginpar{\textbf{[Table~1]}}
Also, Figures~8 and 9 show the snapshots on the $a$-$e$ plane at the end of all EJS and CJS simulations, respectively. 
\marginpar{\textbf{[Figs.~8 and~9]}}
For both EJS and CJS simulations, the total mass and the typical size of planets decrease with increasing 
gas decay time $\tau_{\rm decay}$ because of Type I migration.
When the surface density is large, protoplanets rapidly grow and then spiral into the Sun. 
Therefore, the final total mass $M_{\rm T, final}$ depends on the initial surface density only 
weakly for a certain $\tau_{\rm decay}$.
This trend is particularly evident  when $\tau_{\rm decay} \ge 2$ Myr, 
because the mass depletion due to Type I migration propagates to the outer asteroidal region for large $\tau_{\rm decay}$. 
The number of planets $N_{\rm p}$ is usually 3 or 4.  
In the EJS simulations with $\tau_{\rm decay} = 5$ Myr, however, $N_{\rm p}$ is smaller;
in these simulations, the $\nu_5$ resonance effectively sweeps most of the remnant mass to its current 
location (0.61AU), and a planet with dominant mass ($0.3-0.5$ $M_{\oplus}$) forms there.

We plot the final planets on the plane of semimajor axis versus mass (Fig.~10). 
\marginpar{\textbf{[Fig.~10]}}
The crosses on the figure represent the current terrestrial planets. 
The EJS simulations clearly have better radial mass concentration than the CJS simulations.
In the CJS simulations, the effect of the $\nu_5$ resonance is negligible so that we inevitably have a massive planet 
around 2 AU, quite in contrast to the EJS simulations.
In both EJS and CJS simulations, sizes of outer planets are smaller when the initial surface density in the 
asteroidal region is small, as is the case for small $M_{\rm T}$ or large $p$.
In addition, for the EJS simulations, the radial mass concentration is indirectly affected by the initial 
spatial distribution of planetesimals via their interactions with Jupiter. 
Asteroidal particles reduce $e_{\rm J}$ either by angular momentum exchange at the $\nu_5$ resonance or by energy 
exchange in close encounters with Jupiter; in both cases, the decrease in $e_{\rm J}$ is given as 
$\Delta e_{\rm J} \sim m/(m_{\rm J} e_{\rm J})$, where $m_{\rm J}$ is Jupiter's mass. 
With decreasing $e_{\rm J}$, the effect of the $\nu_5$ resonance decreases (Nagasawa et al., 2005).
If the total mass in the asteroidal region is small due to small $M_{\rm T}$, large $p$, or large $\tau_{\rm decay}$,
the decrease in $e_{\rm J}$ is small (see final $e_{\rm J}$ in Table~1). Such systems can obtain 
high radial mass concentration since Jupiter's perturbation remains effective.    
This accounts for the large difference in the radial mass concentration between cases with 
$M_{\rm T} =5$ and 10 $M_{\oplus}$ in the EJS simulations (Fig.~10).
The $\nu_6$ resonance is also important for the radial mass concentration, and it also works effectively when 
$e_{\rm S}$ is large (we will discuss its effect at Fig.~17 more in detail).
Since Saturn's eccentricity $e_{\rm S}$ is primarily determined by the eccentricity forced by Jupiter, $e_{\rm S}$ is 
usually comparable to or slightly larger than $e_{\rm J}$.

Figure~11 shows the plot of the final planets on the plane of mass versus orbital eccentricity.
\marginpar{\textbf{[Fig.~11]}}
The small orbital eccentricities are well reproduced in the EJS simulations with   
$M_{\rm T} = 5$ $M_{\oplus}$ but not in those with $M_{\rm T} = 10$ $M_{\oplus}$. 
This difference is closely related to the radial mass concentration.
In the case with $M_{\rm T} = 5M_{\oplus}$, the $\nu_5$ resonance effectively sweeps most of the bodies and 
giant impacts between protoplanets complete early while there remains a sufficient amount of gas and planetesimals.
On the other hand, the systems starting with $M_{\rm T} = 10$ $M_{\oplus}$ radially expand even after the secular 
resonance passage and Jupiter continuously perturbs the outer planets.
These planets eventually trigger orbital instability and cause giant impacts after both gas and planetesimals 
significantly deplete.   
Giant impacts tend to occur very late in the CJS simulations. 
However, small asteroidal bodies remain for a long time here as well and tend to damp eccentricities of the planets.
This is why planets in CJS simulations have moderately small eccentricities both for $M_{\rm T} = 5$ and 10 $M_{\oplus}$. 
For the CJS simulations, we found a simple correlation between the mass weighted mean orbital eccentricity, 
$e_m,$ and the mass weighted mean semimajor axis, $a_m$ (Fig.~12).  
\marginpar{\textbf{[Fig.~12]}}
The longer the gas decay time, the larger $a_m$, because mass depletion due to Type I migration propagates outward 
with time. Since Jupiter perturbs outer planets more strongly than inner planets, we have this simple correlation. 
The CJS simulations always produce too large a value for $a_m$, while the EJS simulations well reproduce the current 
value of $a_m$ except for runs with $\tau_{\rm decay} = 5$ Myr where $a_m$ becomes too small.
           
Figure~13 shows angular momenta and impact velocities of all potentially Moon-forming impacts as functions of time.
\marginpar{\textbf{[Fig.~13]}}
The most important result is that giant impacts occur most commonly at $\sim 10$ Myr in the EJS simulations 
whereas at $\sim 100$ Myr in the CJS simulations.
In the EJS simulations, some early giant impacts occur between protoplanets trapped in the $\nu_5$ resonance. 
It is more common, however, that giant impacts occur slightly later when orbital eccentricities of planets 
have sufficiently damped after passage of the resonance so that mutual gravitational attraction is enhanced. 
In the CJS simulations, giant impacts usually occur after gas completely depletes and planetesimals in the 
asteroidal region also significantly (but not completely) deplete. The time of the last Moon-forming impact in each 
simulation is shown in Table~1.
In fact, last giant impacts in some of the EJS simulations occur very late. These systems, however, inevitably have
either large eccentricities or low radial mass concentrations.  

When we see a single simulation in Fig.~13, the impact velocity, $v_{\rm imp}$, is as small as 
the mutual escape velocity $v_{\rm esc}$ at early time, but $v_{\rm imp}/v_{\rm esc}$ increases with time so that  
its value for the last impact is notably larger than unity. 
Some simulations, however, have final impacts with $v_{\rm imp}/v_{\rm esc} \simeq 1$, which is favorable for the 
small iron content of the Moon (Canup, 2004, 2008).
When the orbital separation of two planets, normalized by the mutual Hill radius, slowly decreases due to their growth 
or weak perturbation from other planets, an impact between them with $v_{\rm imp}/v_{\rm esc} \simeq 1$ can
occur after the Jacobi integral in Hill's equation slightly exceeds the potential at the Hill's sphere. 
Orbits of other protoplanets are affected only weakly during the time that two colliding protoplanets interact.
On the other hand, if three or larger number of planets interact strongly, their relative velocity increases to 
the escape velocity. In this case, the typical impact velocity is $v_{\rm imp}/v_{\rm esc} \simeq \sqrt{2}$. 
Further higher $v_{\rm imp}/v_{\rm esc} $ are possible between small protoplanets stirred by larger planets or 
if Jupiter's perturbation is strong. The former type of slow giant impacts usually occur in early stages 
while the latter type of high-velocity giant impacts occur after depletion of both gas and planetesimals (Fig.~13). 
Systems with $S_d$ and $e_m$ as small as those for the current system do not usually experience the latter type of impacts. 
Therefore, not only the small orbital eccentricities of the Earth and Venus but also the small iron content of the Moon
are likely to support the presence of some damping forces at the time of the Moon-forming impact.
      
\subsubsection{Cases with larger eccentricities of giant planets: EEJS and CJSECC simulations}
As shown in Table~1, $e_{\rm J}$ (and $e_{\rm S}$) decreases via interactions with particles in the asteroidal region.
Remnant planetesimals orbiting outside of Jupiter's orbit may also reduce $e_{\rm J}$, although we do not 
include these bodies in our simulations. 
Therefore, the initial $e_{\rm J}$ and $e_{\rm S}$ are likely to have 
been larger than those adopted in the EJS simulations, if their initial 
semimajor axes were (nearly) the same as the present values.
On the other hand, if their initial semimajor axes were those given in the Nice model,
$e_{\rm J} $ and $e_{\rm S} $ might have been nearly zero throughout the terrestrial planet formation, but 
larger initial  $e_{\rm J}$ and $e_{\rm S}$ are not excluded.
The summary of simulations with larger initial $e_{\rm J}$ and $e_{\rm S}$, 
namely the EEJS and CJSECC simulations, is given in Table~2.
\marginpar{\textbf{[Table~2]}} 
The snapshots of planets in the end of simulations in the $a$-$e$ plane are also shown 
in Figs.~14 (the EEJS runs) and 15 (the CJSECC runs). 
\marginpar{\textbf{[Figs.~14 and~15]}} 

In the EEJS simulations, the $\nu_5$ resonance is so strong that $a_m$ usually becomes too small. 
The number of planets is typically one or two, because protoplanets experience large orbital excursions.
None of the simulations have $S_c$ as large as the present value, because the semimajor axis of the innermost planet 
is usually too small. The innermost planet is pushed toward the Sun by outer planets which are 
swept by the $\nu_5$ resonance. The potentially Moon-forming impacts usually occur too early and 
impact velocities are too high (Fig.~16).
\marginpar{\textbf{[Fig.~16]}} 
Clearly, the initial $e_{\rm J}$ of our EEJS simulations is too high. If $e_{\rm J}$ is somewhat smaller but still 
larger than the present value, the $\nu_5$ resonance may work much better (Thommes et al., 2008).

The initial orbital eccentricities of Jupiter and Saturn in the CJSECC simulations are similar to those in the 
EJS simulations. The orbital separation between Jupiter and Saturn in the CJSECC simulations is narrower than 
that in the EJS simulations. This makes the $\nu_5$ and $\nu_6$ resonances in the CJSECC simulations 
more distant from the Sun than those in the EJS simulations.
Because of this, the averaged $a_m$ and $t_{\rm imp}$ in the CJSECC simulations become larger than those in the 
EJS simulations (Tables~1 and 2).
Figure~17 shows a comparison between a CJSECC and an EJS simulation with the same parameters except for the orbits of 
Jupiter and Saturn.
\marginpar{\textbf{[Fig.~17]}} 
The $\nu_5$ resonance sweeps planetesimals effectively at 3 Myr but no longer works at 5 Myr in both cases.
In the EJS simulation, however, the $\nu_6$ resonance ($\sim$ 2 AU) still sweeps planetesimals toward the 
terrestrial region. On the other hand, the location of the $\nu_6$ resonance in the CJSECC simulation is too 
far ($\sim$ 3 AU) to contribute to radial mass concentration. 
In the CJSECC simulation, accretion of outer bodies, which survived the sweeping by the $\nu_5$ resonance, 
proceeds slowly and this makes the final giant impact late.
Surprisingly, 7 out of 10 CJSECC simulations in which potentially Moon-forming impacts occur 
satisfy the time of the last giant impact suggested from isotope records (Table~2). 
However, $v_{\rm imp}/v_{\rm esc}$'s of most of these late impacts are too high (Fig.~16) and also $S_c$'s of these 
systems are too small (Table~2). 

As we have seen, simulation outcomes strongly depend on Jupiter's eccentricity, $e_{\rm J}$. 
Through interactions with small bodies, $e_{\rm J}$ decreases. Clearly, the effect of Jupiter's perturbation is more 
important if $e_{\rm J}$ in the late stage is larger regardless of initial values of $e_{\rm J}$. 
In Fig.~18, we plot the angular momentum deficit $S_d$, the radial mass concentration parameter $S_c/a_m^2$, 
and the time of the last one of potentially Moon-forming impacts $t_{\rm imp}$ versus the value of $e_{\rm J}$ at the 
end of a simulation, $e_{\rm J, final}$ (averaged over a few Myr).
\marginpar{\textbf{[Fig.~18]}} 
We find that $S_c/a_m^2$ more appropriately represents the radial mass confinement than does $S_c$ 
because systems consisting of small planets with large $a$ tend to obtain large $S_c$ while $b_{min}$ is large 
(particularly in the CJS simulations). 
Simple linear $\chi^2$ fits suggest that $S_c/a_m^2$ and $t_{\rm imp}$ increases/decreases with $e_{\rm J, final}$,
although the scatter of the simulation data are very large.
The problem is that $S_c/a_m^2$ can be as large as the present value when $e_{\rm J, final}$ is large
whereas $t_{\rm imp}$ can be as large as that suggested from isotope records only when $e_{\rm J, final}$
is small enough.
For $S_d$, we fit the data with a quartic function as two competing effects are expected: 
large $e_{\rm J}$ means stronger perturbation while quick accretion due to large $e_{\rm J}$ means
stronger damping forces due to remnant gas and planetesimals. 
We find that $S_d$ is nearly flat for small $e_{\rm J, final}$ and increases with increasing $e_{\rm J,final}$ 
when $e_{\rm J,final}$ is larger than the present value. 
In an ensemble average sense, $S_d$ is usually much larger than the current value, although some simulations have 
$S_d$ consistent or even smaller than the present day value.
The dependence of all three properties on $e_{\rm J, final}$ is stronger for the CJS+CJSECC simulations than
for the EJS+EEJS simulations. It is unclear to us, however, whether this is really physically meaningful because 
fits for the CJS+CJSECC simulations are not very reliable with only a few points at large $e_{\rm J}$.

\section{Discussion}

None of our simulations succeeded in satisfying all of the three constraints listed in Sec.~3.2.   
The most serious issue we found is a trade-off between the radial mass concentration parameter, $S_c/a_m^2$, 
and the time of the potentially Moon-forming impact, $t_{\rm imp}$ (the upper panel of Fig.~19).
\marginpar{\textbf{[Fig.~19]}}  
The large $S_c/a_m^2$ in the current system is due to the small masses of Mars and Mercury and the small orbital separation between 
the Earth and Venus.  
It is hard to satisfy both of these conditions in a system with large $t_{\rm imp}$.
Since the evolution time scale is longer in the outer regions, giant impacts can occur late when a large amount of mass stays in the 
outer region for a long time.  Such a system usually has outer planets more massive than Mars. 
Also, if a planetary system slowly evolves with weak perturbations from Jupiter (e.g., the CJS simulations),
the orbital separation between neighboring planets normalized by the mutual Hill radius inevitably 
becomes larger than that between the Earth and Venus (26.3) owing to the mutual orbital repulsion. 
If $e_{\rm J}$ is as large as the present value or larger, $S_c/a_m^2$ can be as large as today's value because Jupiter tends 
to radially compress a system. However, accretion completes too rapidly in such a case.

There is also a trade-off between the angular momentum deficit $S_d$ and  $t_{\rm imp}$ (the lower panel of Fig.~19). 
Namely, in order to obtain small $S_d$,
a final large impact needs to occur while some damping forces work effectively.
Otherwise, $S_d$ ends up with much larger than the current value.
This trend is particularly clear for the EJS and CJS simulations  (filled and open circles).
Small $S_d$ even without any damping forces seems to be possible but such a case is statistically rare.

Our failure indicates that there may be some missing physics in our simulations or flaws with the assumptions that we have adopted. 
In the following, we suggest three possible scenarios/solutions for future studies.
 
In the case with an eccentric Jupiter, a possible solution is that $\tau_{\rm decay}$ is much longer and Type I migration is 
much slower than we assumed.  This situation actually corresponds to that adopted in the original dynamical shake up model of
Nagasawa et al. (2005) and Thommes et al. (2008). In their simulations ($\tau_{\rm decay} = 3$ and 5 Myr),  
the giant impacts usually occur at $t \simeq 4-5$ $\tau_{\rm decay}$ when the $\nu_5$ resonance reaches near 1 AU.
Therefore, $\tau_{\rm decay} > 10$ Myr is required for the late Moon-forming impact indicated from isotope records
($t_{\rm imp} > $ 50 Myr; Touboul et al. 2007).  
If $\tau_{\rm decay} = 10$ Myr, the gas density is still $\sim 1 \%$ of the minimum mass solar nebular at $50$ Myr,
and observations suggest that such a gas disk is exceptionally long-living (Pascucci et al. 2006). 
Slightly smaller $t_{\rm imp} (\simeq $ 30 Myr) is not excluded from the terrestrial W/Hf records only (Jacobsen et al., 2009),
but this still requires a long-surviving disk ($\tau_{\rm decay} \sim 6$ Myr).
Recent theories indicate significant changes in Type I migration rate due to the non-isothermal effect 
(Baruteau and Masset, 2008; Paardekooper and Mellema, 2008) and the non-linear effect of the co-rotation resonance 
(Masset et al., 2006; Paardekooper and Paparoizou, 2009).
These mechanisms cause slow down of inward migration or
even outward migration if the radial entropy gradient is negative for the former mechanism 
or if the radial surface density gradient is nearly flat or positive for the latter one.
Therefore, if a local density maximum exists, planetary migration slows down or protoplanets even tend to migrate toward 
this point (Kretke et al., 2009).  A local density maximum can exist at the boundary between 
the optically thin inner region and the optically thick outer region (i.e., the inner boundary of the so-called dead-zone; Kretke and Lin, 2007; Brauer et al., 2008).  Protoplanets might have survived in the protosolar nebular for a long time with such mechanisms.

An alternative possibility for the case with an eccentric Jupiter is that
the nebular gas inside of Jupiter might have dissipated very rapidly after the formation of Jupiter. 
This situation corresponds to previous $N$-body simulations for the late stages without gas 
(Chambers, 2001; O'Brien et al., 2006;  Raymond et al., 2009).
Without gas, $t_{\rm imp}$ tends to be larger than those in our simulations, because the effect of $\nu_5$ is less signifiant. 
Non-uniform gas dissipation models, such as inside-out and gap-opening models discussed in Nagasawa et al. (2000),
probably have similar outcomes, because $\nu_5$ passes a certain location after the gas dissipates there.
Raymond et al. (2009) showed that some of their EEJS simulations nearly satisfy all the three constraints listed in Sec.~3.2.  
In fact, these simulations satisfy the small mass of Mars, but not the small orbital separation between the Earth and Venus. This is why
the values of $S_c$ in all of their simulations are smaller than the current value. This problem might be resolved, however, 
if they adopt even higher resolution with smaller planetesimals (e.g., Morishima et al., 2008) and inelastic rebounds for high-speed or grazing collisions 
(Agnor and Asphau, 2004; Asphau et al., 2006). Because these effects are likely to enhance the damping effects and reduce 
the orbital excursions of planets, the orbital separation between neighboring planets will probably be smaller.
Very quick or non-uniform dissipation of gas seems to be favorable, at least, in the asteroidal region in the case with large $e_{\rm J}$.
As we have shown, the $\nu_5$ resonance is so strong in the asteroidal region, except the CJS simulations, 
that the radial compositional structure in the asteroidal region is destroyed even for $\tau_{\rm decay} = 1$ Myr.
The gas dissipation of the order of 0.1 Myr or shorter seems to be necessary for the uniform dissipation model (Nagasawa et al., 2000).
Jupiter might have opened a wide gap around its orbit, extending to the entire asteroidal region immediately 
after its formation (however, $\tau_{\rm decay} = 1$ Myr or larger may be still possible if $e_{\rm J}$ is small, when 
$\nu_5$ passes the asteroid region, and increases later). 
 
In the case of a circularly orbiting Jupiter, the initial planetesimals needed to be 
radially highly concentrated in order to satisfy a large $S_c$. In other words, the depletion of mass in the asteroidal region needed to 
occur before the formation of  planetesimals.  Such a concentration of planetesimals may be possible if 
planetesimal formation occurs in radially limited regions after the increase of the surface density due to dust migration 
(Youdin and Shu, 2002; Youdin and Chiang, 2004; Kretke and Lin, 2007; Brauer et al., 2008). 
Morishima et al. (2008) performed $N$-body simulations starting from highly concentrated disks in the absence of gas, and found that some of the resulting planetary systems 
were similar to the current solar system.
The problem of their simulations is that accretion usually completes very quickly ( $< 10$ Myr), as well as our EJS simulations. 
Nevertheless, one simulation (run 1 of Morishima et al., 2008)
has a very late giant impact.  This system unfortunately has slightly too large an orbital separation between the two largest planets.
Hansen (2009) also made simulations similar to Morishima et al. (2008) and found the mean time of last giant impacts to be 45 Myr.
However, the lower limit to the impactor mass (0.02 $M_{\oplus}$) used for his definition of giant impacts is probably too small for formation the Moon.  
Further explorations for planetary accretion starting from radially concentrated disks are necessary.

\section{Conclusions} 
We have performed $N$-body simulations which follow the growth and accretion histories of the terrestrial planets, including the effects of the giant planets and the interaction with gas in laminar disks.
We vary initial total mass and spatial distribution of planetesimals, the time scale of dissipation of a gas disk,
and the orbits of Jupiter and Saturn. It is a remarkable success of numerical simulations, that planetary systems with global properties similar to the terrestrial planets can be obtained starting with a proto-planetary disk. We are now at the stage where fine details of the models can be compared with solar system constraints.
In order to retain a final total mass as massive as 
the present total mass ($\sim$ 2 Earth mass) in the presence of standard Type I migration, the time scale of dissipation of gas needs to be 1-2 Myr.

Using the results from a large number of simulations we compared the evolutionary histories of the resulting planetary systems with the 
following three properties of the terrestrial planets:
(1) the high radial mass concentration, (2) the small orbital eccentricities and (3) the very late Moon-forming impact ($\sim 100$ Myr).
The orbital eccentricities of the planets can be as small as their present values if giant impacts finish whilst there is still remnant 
gas or planetesimals, as expected.  
The most serious issue we found was a trade off between the radial mass concentration and the late Moon-forming impact.
If the orbital eccentricities of Jupiter are as large as the present value, most of bodies in the asteroidal region are swept up to the terrestrial region 
owing to inward migration of the secular resonance. Although this mechanism helps planetary systems to have high radial concentrations in the end,  giant impacts between protoplanets most commonly occur too early ($\sim$10 Myr).   
On the other hand,  if the orbital eccentricity of Jupiter is close to zero as 
suggested in the Nice model, the effect of the secular resonance is negligible 
and a large amount of mass stays for a long period of time 
in the asteroidal region. Although giant impacts usually occur around 100 Myr, 
we inevitably have an Earth size planet at around 2 AU in this case. 
It is very difficult to obtain spatially concentrated terrestrial planets 
with very late giant impacts, as long as we include all the effects of gas and assume initial disks similar to the minimum mass solar nebular.

\section*{Acknowledgements}
We appreciate comments on our manuscript from two anonymous reviewers and the editor, Alessandro Morbidelli.
R.M. thanks Sei-ichiro Watanabe for his hospitality during the visit of R.M. and fruitful discussions.
Simulations were performed on the Zbox2 and Zbox3 supercomputers at the University of Zurich.

\section*{References}
\begin{description}
\item 
Adachi, I., Hayashi, C., Nakazawa, K. 1976.
The gas drag effect on the elliptical motion of a solid body in the primordial solar nebula.	
Prog. Theoret. Phys. 56, 1756--1771.

\item 
Agnor, C.B., Asphau, E. 2004.
Accretion efficiency during planetary collisions.
Astrophys. J. 613, L157--L160.
 
\item 
Agnor, C.B., Ward, W.R. 2002.
Damping of terrestrial-planet eccentricities by density-wave interactions with a remnant gas disk.
Astrophys. J. 567, 579--589.

\item
Agnor, C.B., Canup, R.M., Levison, H.F. 1999.
On the character and consequences of large impacts in the late stage of terrestrial planet formation.
Icarus 142, 219--237.

\item
All\'{e}gre, C.J., Manh\`{e}s, G., G\"{o}pel, C.  2008.
The major differentiation of the Earth at  $\sim$4.45 Ga.
Earth Planet. Sci. L. 267, 386--398.   

\item
Asphau, E., Agnor, C.B., Quentin, W. 2006.
Hit-and-run planetary collisions.
Nature 439, 7073, 155--160.

\item
Baruteau, C., Masset, F. 2008.
On the corotation torque in a radiatively inefficient disk.
Astrophys. J. 672, 1054--1067.
			
\item
Bodenheimer,  P., Pollack, J.B. 1986.
Calculations of the accretion and evolution of giant planets: The effects of solid cores.
Icarus 67, 391--408. 

\item
Boley, A.C. 2009.
The two modes of gas giant planet formation.
Astrophys. J. 695, L53--L57.

\item
Brauer, F., Henning, Th., Dullemond, C.P. 2008.
Planetesimal formation near the snow line in MRI-driven turbulent protoplanetary disks.
Astrophys. Astron. 487, L1--L4.

\item
Cameron, A.G.W., Benz, W. 1991.
The origin of the moon and the single impact hypothesis. IV.
Icarus 92, 204--216.

\item
Canup, R.M. 2004.
Simulations of a late lunar-forming impact.
Icarus 168, 433--456.

\item
Canup, R.M. 2008.
Lunar-forming collisions with pre-impact rotation.
Icarus 196, 518--538.

\item
Canup, R.M., Ward, W.R., Cameron, A.G.W. 2001.
A scaling relationship for satellite-forming impacts.
Icarus 150, 288--296.

\item 
Chambers, J.E.  1999.
A hybrid symplectic integrator that permits close encounters between massive bodies.
Mon. Not. R. Astron. Soc. 304, 793--799.

\item
Chambers, J.E.  2001.
Making more terrestrial planets.
Icarus, 152, 205--224.

\item
Chambers, J.E., Cassen, P. 2002.
The effects of nebula surface density profile and giant-planet eccentricities on planetary accretion in the inner solar system.
Meteo. Planet. Sci. 37, 1523--1540.

\item
Chambers, J.E., Wetherill, G.W.  1998.
Making the terrestrial planets: $N$-body integrations of planetary embryos in three dimensions.
Icarus 136, 304--327

\item
Chambers, J.E., Wetherill, G.W., Boss, A.P. 1996.
The stability of multi-planet systems.
Icarus 119, 261--268.

\item
Cresswell, P., Dirksen, G., Kley, W., Nelson, R.P. 2007.
On the evolution of eccentric and inclined protoplanets embedded in protoplanetary disks.
Astron. Astrophys. 473, 329--342.

\item
Crida, A., Morbidelli, A. 2007.
Cavity opening by a giant planet in a protoplanetary disc and effects on planetary migration.
Mon. Not. R. Astron. Soc. 377, 1324--1336.

\item
Daisaka, J.K., Tanaka, H., Ida, S. 2006.
Orbital evolution and accretion of protoplanets tidally interacting with a gas disk. II. Solid surface density evolution with Type-I migration.
Icarus 185, 492--507.

\item
D'Angelo, G., Lubow, S.H., Bate, M.R. 2006.
Evolution of giant planets in eccentric disks.
Astrophys. J. 652, 1698--1714.	
	
\item
Duncan, M.J., Levision, H.F.,  Lee, M.H. 1998.
A multiple time step symplectic algorithm for integrating close encounters.
Astron. J. 116, 2067--2077.

\item
Goldreich, P.,  Sari, R. 2003.
Eccentricity evolution for planets in gaseous disks.
Astrophys. J. 585, 1024--1037.

\item 
Goldreich, P., Tremaine, S. 1980.
Disk-satellite interactions.
Astrophys. J. 241, 425--441.
	
\item
Goldreich, P., Ward, W.R. 1973.
The formation of planetesimals.
Astrophys. J. 183, 1051-1062.

\item
Gomes, R., Levison, H.F., Tsiganis, K., Morbidelli, A. 2005.
Origin of the cataclysmic Late Heavy Bombardment period of the terrestrial planets.
Nature 435, 466--469.
	
\item
Gradie, J., Tedesco, E. 1982.
Compositional structure of the asteroid belt.
Science 216, 1405--1407.

\item 
Hansen, B. 2009.
Formation of the terrestrial planets from a narrow annulus.
Astrophys. J.,  703, 1131-1140.

\item
Hayashi, C. 1981.
Structure of the solar nebula, growth and decay of magnetic fields and effects of magnetic and turbulent viscosities on the nebula.
Suppl. Prog. Theoret. Phys. 70,  35--53.

\item
Heppenheimer, T.A. 1980.
Secular resonances and the origin of eccentricities of Mars and the asteroids.
Icarus 41, 76-88.

\item
Ida, S., Lin, D.N.C. 2008.
Toward a deterministic model of planetary formation. IV. Effects of Type I migration.
Astrophys. J. 673, 487--501.

\item
Ida, S., Canup, R.M., Stewart, G.R. 1997.
Lunar accretion from an impact-generated disk.
Nature 389, 353--357.

\item
Ikoma, M., Nakazawa, K., Emori, H. 2000.
Formation of giant planets: dependences on core accretion rate and grain opacity.	
Astrophys. J. 537, 1013--1025.	
	
\item
Inaba, S., Tanaka, H., Nakazawa, K., Wetherill, G.W., Kokubo, E. 2001.
High-accuracy statistical simulation of planetary accretion: II. Comparison with $N$-body simulations. 
Icarus 149, 235--250.

\item
Ito, T., Tanikawa, K. 1999.
Stability and instability of the terrestrial protoplanet system and their possible roles in the final stage of planet formation.
Icarus, 139, 336--349.

\item
Jacobsen, S.B., Remo, J.L., Petaev, M.I., Sasselov, D.D. 2009.
Hf-W chronometry and the timing of the giant Moon-forming impact on Earth.
40th Lunar and Planetary Science Conference abstract, A2054. 

\item
Kokubo, E.,  Ida, S.  1996.
On runaway growth of planetesimals.
Icarus 123, 180--191.

\item
Kokubo, E.,  Ida, S. 1998.
Oligarchic growth of protoplanets.
Icarus 131, 171--178.

\item 
Kokubo, E., Ida, S. 2000.  
Formation of protoplanets from planetesimals in the Solar nebula.
Icarus 143, 15--27.

\item
Kokubo, E., Ida, S. 2002.  
Formation of protoplanet systems and diversity of planetary systems.
Astrophys. J. 581, 666--680.

\item 
Kokubo, E., Kominami, J., Ida, S. 2006.  
Formation of terrestrial planets from protoplanets. I. Statistics of basic dynamical properties.
Astrophys. J. 642, 1131--1139.

\item
Kominami, J., Ida, S. 2002.
The effect of tidal interaction with a gas disk on formation of terrestrial planets.
Icarus 157, 43--56.

\item
Kominami, J.,  Ida, S. 2004.  
Formation of terrestrial planets in a dissipating gas disk with Jupiter and Saturn.
Icarus 167, 231--243.

\item
Kretke, K.A., Lin, D.N.C. 2007.
Grain retention and formation of planetesimals near the snow line in MRI-driven turbulent protoplanetary disks.
Astrophys. J. 664, L55--L58.

\item
Kretke, K.A., Lin, D.N.C., Garaud, P., Turner, N.J. 2009.
Assembling the building blocks of giant planets around intermediate-mass stars.
Astrophys. J. 690, 407--415.

\item
Laskar, J. 1997.
Large scale chaos and the spacing of the inner planets.
Astron. Astrophys. 317, L75--L78.

\item
Lissauer, J.J., Hubickyj, O., D'Angelo, G., Bodenheimer, P. 2009.
Models of Jupiter's growth incorporating thermal and hydrodynamic constraints.
Icarus 199, 338--350.

\item
Masset, F.S., D'Angelo, G., Kley, W. 2006.
On the migration of protogiant solid cores.
Astrophys. J. 652, 730--745.

\item
McNeil, D., Duncan, M., Levison, H.F. 2005.
Effects of Type I migration on terrestrial planet formation.
Astron. J. 130, 2884--2899.

\item
Morbidelli, A., Crida, A. 2007.
The dynamics of Jupiter and Saturn in the gaseous protoplanetary disk.
Icarus 191, 158--171.	
	
\item 
Morishima, R., Schmidt, M.W., Stadel, J., Moore, B. 2008.
Formation and accretion history of terrestrial planets from runaway growth through to late time: Implications for orbital eccentricity.
Astrophys. J. 685, 1247--1261.


\item
Muzerolle, J., Calvet, N., Brice\~{n}o, C., Hartmann, L., Hillenbrand, L. 2000.
Disk accretion in the 10 Myr old T Tauri stars TW Hydrae and Hen 3-600A.
Astrophys. J. 535, L47--L50. 

\item
Nagasawa, M., Lin, D.N.C., Thommes, E.W. 2005.
Dynamical shake-up of planetary systems. I. Embryo trapping and induced collisions by 
the sweeping secular resonance and embryo-disk tidal interaction.
Astrophys. J. 635, 578--598.

\item
Nagasawa, M., Tanaka, H., Ida, S. 2000.   
Orbital evolution of asteroids during depletion of the solar nebula.
Astron. J. 119, 1480--1497.

\item
O'Brien, D.P., Morbidelli, A., Levison, H.F. 2006.
Terrestrial planet formation with strong dynamical friction.
Icarus 184, 39-- 58.

\item
Ogihara, M., Ida, S., Morbidelli, A. 2007.
Accretion of terrestrial planets from oligarchs in a turbulent disk.
Icarus 188, 522--534.

\item 
Oglvie, G.I., Lubow, S.H. 2003.
Saturation of the corotation resonance in a gaseous disk.
Astrophys. J. 587, 398--406. 

\item
Paardekooper, S.-J., Mellema, G. 2008.
Growing and moving low-mass planets in non-isothermal disks.
Astron. Astrophys. 478, 245--266.

\item
Paardekooper, S.-J., Papaloizou, J.C.B. 2009.
On corotation torques, horseshoe drag and the possibility of sustained stalled or outward protoplanetary migration.
Mon. Not. R. Astron. Soc. 394, 2283--2296.

\item
Papaloizou, J.C.B., Larwood, J.D. 2000.
On the orbital evolution and growth of protoplanets embedded in a gaseous disc.
Mon. Not. R. Astron. Soc. 315, 823--833.
 	
\item
Pascucci, I., and 19 co-authors, 2006.
Formation and evolution of planetary systems: Upper limits to the gas mass in disks around Sun-like stars.
Astrophys. J. 651, 1177--1193.

\item
Quinn, T.,  Tremaine, S., Duncan, M. 1991.
A three million year integration of the earth's orbit.
Astron. J. 101, 2287--2305.

\item
Raymond, S.N., O'Brien, D.P., Morbidelli, A., Kaib, N.A. 2009.
Building the terrestrial planets: Constrained accretion in the inner solar system.
Icarus 203, 644--662.

\item
Raymond, S.N., Quinn, T., Lunine, J.I. 2004. 
Making other earths: dynamical simulations of terrestrial planet formation and water delivery.
Icarus 168, 1--17. 

\item
Raymond, S.N., Quinn, T., Lunine, J.I. 2006.
High-resolution simulations of the final assembly of Earth-like planets I. Terrestrial accretion and dynamics.
Icarus 183, 265--282. 

\item
Richardson, D.C., Quinn, T., Stadel, J., Lake, G. 2000. 
Direct large-scale $N$-body simulations of planetesimal dynamics.
Icarus 143, 45--59. 

\item
Stadel, J. 2001.
Cosmological $N$-body simulations and their analysis.
PhD dissertation, Univ. of Washington, Seattle.

\item
Strom, R.G., Malhotra, R., Ito, T., Yoshida, F., Kring, D.A. 2005.
The origin of planetary impactors in the inner solar system.
Science 309, 1847--1850.

\item 
Tanaka, H., Ward, W.R. 2004.
Three-dimensional interaction between a planet and an isothermal gaseous disk. II. Eccentricity waves and bending waves.
Astrophys. J. 602, 388--395.

\item 
Tanaka, H., Takeuchi, T., Ward, W.R. 2002.
Three-dimensional interaction between a planet and an isothermal gaseous disk. 
I. Corotation and Lindblad torques and planet migration.
Astrophys. J. 565, 1257--1274.
	
\item 
Touboul, M., Kleine, T., Bourdon, B., Palme, H., Wieler, R. 2007.
Late formation and prolonged differentiation of the Moon inferred from W isotopes in lunar metals.
Nature 450, 1206--1209.
		
\item
Thommes, E., Nagasawa, M., Lin, D.N.C. 2008.
Dynamical shake-up of planetary systems. II. $N$-body simulations of solar system 
terrestrial planet formation induced by secular resonance sweeping.
Astrophys. J. 676, 728--739.
	
\item		
Tsiganis, K., Gomes, R., Morbidelli, R., Levision, H.F. 2005. 
Origin of the orbital architecture of the giant planets of the Solar System. 
Nature 435, 459--461.		
		
\item
Yoshinaga, K., Kokubo, E., Makino, J. 1999.
The stability of protoplanet systems.
Icarus 139, 328--335.

\item
Youdin, A.N., Chiang, E.I. 2004.
Particle pileups and planetesimal formation.
Astrophys. J. 601, 1109--1119.

\item
Youdin, A.N., Shu, F.H. 2002.
Planetesimal formation by gravitational instability.
Astrophys. J. 580, 494--505.

\item
Wadsley, J.W., Stadel, J., Quinn, T. 2004.
Gasoline: a flexible, parallel implementation of TreeSPH.
New Astronomy 9, 137--158.

\item
Ward, W.R. 1981.
Solar nebula dispersal and the stability of the planetary system. I - Scanning secular resonance theory.
Icarus 47, 234-264.

\item
Ward, W.R. 1986.
Density waves in the solar nebula - Differential Lindblad torque.
Icarus 67, 164--180.

\item
Ward, W.R. 1997. 	
Protoplanet migration by nebula tides.
Icarus 126, 261--281. 
		
\end{description}

\clearpage

\begin{table}
\begin{center}
\footnotesize
\begin{tabular}{|c|cccccccccccc|} \hline
 JS orbits   &  $\tau_{\rm decay}$ & $p$ & $M_{\rm T}$& $N_{\rm p}$& $M_{\rm T, final}$  & $S_c$ & $b_{\rm min}$ & $a_m$&$ e_m$ & $S_d$ & $t_{\rm imp}$ & $e_{\rm J, final}$\\ 
                    &    (Myr)                       &    & ($M_{\rm E}$)&                  &($M_{\rm E}$)           &             &                           &    (AU)                      &                 &              &        (Myr)         &       \\ \hline
  EJS & 1 &  1 & 10 & {\bf 4} & 4.30          & 13.8           &  {\bf 30.1} &  1.15         &  0.098         & 0.0074          &  {\bf 85.6}  & 0.006 \\
           & 1 &  1 &  5  & {\bf 3} & 3.08          & 29.3           & 39.8        &  {\bf 0.94}   & {\bf 0.063} & {\bf 0.0032}   & 12.1   & 0.021 \\
           & 1 &  2 & 10 & {\bf 3} & {\bf 2.44} & 31.0           & 40.6         & {\bf 0.97}   & 0.078         & 0.0058            & 27.5           & 0.032 \\
           & 1 &  2 &  5  & {\bf 3} & {\bf 1.93}  & {\bf 78.3}   & {\bf 33.7}  & {\bf 0.95}   & {\bf 0.028} & {\bf 0.0013}   & 13.5          & 0.026 \\
           & 2 &  1 &10  & {\bf 5} & {\bf 1.97}  & 28.1          & {\bf 33.5}   & 1.19           & 0.114         & 0.0109           & 32.8   & 0.002 \\
           & 2 &  1 &  5  & {\bf 4} & {\bf 2.28}  & 17.7          & {\bf 34.7}   & {\bf 0.82}   & {\bf 0.033} & {\bf 0.0010}  & 308.4        & 0.028\\
           & 2 &  2 &10  & {\bf 3} & {\bf 1.49}  & 26.5          & {\bf 34.0}   & {\bf 0.93}   & 0.137          & 0.0119          & {\bf 83.3}   & 0.025 \\
           & 2 &  2 &  5  & {\bf 3} & {\bf 1.18}  & {\bf 79.9}   &42.2          & {\bf 0.97}   & {\bf 0.036} & {\bf 0.0016}  & 13.3         & 0.035 \\
           & 3 &  1 &10  & {\bf 4} & 0.94          & {\bf 69.8}  & {\bf 23.8}  & 1.19           & {\bf 0.055} & 0.0044          & 0.6           & 0.022 \\
           & 3 &  1 &  5  & {\bf 3} & {\bf 1.33}  & 34.5         & 47.2           & {\bf 0.99}   & 0.081          & 0.0145         & 327.4      & 0.030 \\
           & 3 &  2 &10  & 2         & 0.69         & 18.1          & 111.5        & 0.65            & {\bf 0.016}  & {\bf 0.0004}   & --             & 0.040 \\
           & 3 &  2 &  5  & {\bf 3} & 0.78         & {\bf 74.9}  & 42.1           & 1.13            & {\bf 0.064}  & 0.0048           &  --           & 0.035 \\
           & 5 &  1 & 10 & {\bf 3} & 0.46         & 25.8          & 83.1           & {\bf 0.88}    & 0.332         & 0.0599           &  --            & 0.036 \\
           & 5 &  1 &  5  & 1          & 0.51         &  --             & --                 &  0.68            & 0.285         & 0.0446          &  --             & 0.031 \\   
           & 5 &  2 & 10 & 2          & 0.31        & {\bf 70.2} &101.9           &  0.74            & 0.107         & 0.0062          &  --             & 0.038  \\
           & 5 & 2  & 5   & 1          & 0.40       &  --               & --                  & 0.68            & 0.143         & 0.0113          & --              & 0.042 \\ \hline
CJS   & 1 & 1  & 10 & {\bf 4} & 4.32         & 10.9          &35.9              & 1.11           & {\bf 0.058} & {\bf 0.0025}   &{\bf 109.7}         & 0.004 \\
           & 1 & 1  & 5   & {\bf 4} & {\bf 2.62}  & 17.1         &38.3              & {\bf 1.00}   & {\bf 0.064} & {\bf 0.0030}   & {\bf 85.5}   & 0.004 \\  
           & 1 & 2  & 10 & {\bf 4} & {\bf 2.45} & 17.5          &{\bf 34.5}      & 1.35           & 0.083         &  0.0054         & {\bf 117.2} & 0.004 \\ 
           & 1 & 2  & 5   & {\bf 4} & {\bf 2.02}  & 11.7          &43.3             &  1.14          & {\bf 0.071} & 0.0055           & {\bf 139.5}  & 0.003 \\    
           & 2 & 1  & 10 & {\bf 4} & {\bf 2.50}  & 13.0         &38.1             & 1.16            & {\bf 0.073} & 0.0134           & 388.8         & 0.007 \\ 
           & 2 & 1  & 5   & {\bf 4} & {\bf 1.82}  & 33.7          &35.4             & 1.22           &  {\bf 0.045} & {\bf 0.0016}  & 46.1           & 0.004 \\
           & 2 & 2  &10  & 2         & {\bf 1.57}  & {\bf 54.9}  &51.9            & 1.26            &  0.077         & 0.0047          & 184.4       & 0.004 \\
           & 2 & 2  & 5   & {\bf 3} &  0.98         & {\bf 72.3}  &55.9            & 1.49             &  0.100         & 0.0115          &189.5         & 0.003 \\
          & 3  & 1  & 10 & {\bf 3} & {\bf 1.22}  & {\bf 55.7}  &50.5            & 1.46             &  {\bf 0.071} & 0.0081          &{\bf 122.7}  & 0.004 \\
          & 3  & 1  & 5   &  {\bf 4} & {\bf 1.49} & 13.5           &35.8           &  1.06            &  {\bf 0.050} & {\bf 0.0025}  & --                & 0.003 \\
          & 3  & 2  & 10 &  {\bf 3} & 0.93         & 43.1          &52.5            & 1.39             &  0.191        &  0.0306         & 506.2 & 0.005 \\
          & 3  & 2  & 5   &  2         & 0.75          & {\bf 70.4}  &59.7           & 1.63             &  0.095         &  0.0094         & --         & 0.003 \\ 
          & 5  & 1  & 10 &  {\bf 3} & 0.73         & {\bf 52.3}  &56.1            & 1.59             &  0.080         &  0.0098        & --         & 0.004 \\  
          & 5  & 1  & 5   &  {\bf 4} & 0.81         & 19.7          &59.1           & 1.21               &  {\bf 0.074} &  0.0151         & --         & 0.004 \\ 
          & 5  & 2  & 10 &  {\bf 3} & 0.55         & 16.7         &51.7           & 1.49                &  {\bf 0.069} &  0.0167         & --          & 0.003 \\ 
          & 5  & 2  & 5   &  {\bf 4} & 0.53        & 32.0          &61.7           & 1.50                & 0.117          &  0.0159         & --          & 0.003 \\ \hline
  SS   &     &     &       & 4 &  1.98               & 89.9          & 26.3          &  0.90               & 0.038          &  0.0018           & 50--150 & 0.048 \\ \hline          
\end{tabular}
\caption{Simulation parameters and outcomes for the EJS and CJS simulations. 
Table columns are the orbits of Jupiter and Saturn,  the gas decay time scale $\tau_{\rm decay}$,  the power-law index for the radial gradient of 
the initial surface density of planetesimals $p$,  the initial total mass of planetesimals $M_{\rm T}$, the number of planets $N_{\rm p}$,  the final total mass of 
planets $M_{\rm T, final}$,  the radial mass concentration statistics $S_c$, the mass weighted mean semimajor axis $a_m$, the minimum 
orbital separation between neighboring planets $b_{\rm min}$
normalized by the mutual Hill radius, the mass weighted mean eccentricity $e_m$, the angular momentum deficit $S_d$, 
the time of the last potentially Moon-forming impacts $t_{\rm imp}$,  and 
the final orbital eccentricity of Jupiter $e_{\rm J, final}$. 
Any orbital elements averaged over a few Myr are used. Each value is boldly marked when $3 \le N_{\rm p} \le 5$, 
$1.0 \le M_{\rm T, final}/M_{\oplus} \le 3.0$, $S_c > 45.0$, 0.75 AU $< a_m<$ 1.05 AU,  $b_{\rm min} < 35$, $e_m < 0.076$, $S_d < 0.0036$, 
or 50 Myr $< t_{\rm imp}<$ 150 Myr. SS represents the solar system.}
\end{center}
\end{table}

\clearpage

\begin{table}
\begin{center}
\footnotesize
\begin{tabular}{|c|cccccccccccc|} \hline
 JS orbits   &  $\tau_{\rm decay}$ & $p$ & $M_{\rm T}$& $N_{\rm p}$& $M_{\rm T, final}$  & $S_c$ & $b_{\rm min}$ & $a_m$&$ e_m$ & $S_d$ & $t_{\rm imp}$ & $e_{\rm J, final}$\\ 
                    &    (Myr)                       &    & ($M_{\rm E}$)&                  &($M_{\rm E}$)           &             &                           &    (AU)                      &                 &              &        (Myr)         &       \\ \hline
EEJS & 1 &  1 & 10 &  {\bf 3} & 3.70        & 16.6         & 38.1             & 0.72           & 0.149          &  0.0219          & 180.0      & 0.054\\
           & 1 &  1 &  5  &  {\bf 4} & 3.41         & 22.6        & {\bf 30.8}    & {\bf 0.85}   &  {\bf 0.053} &  {\bf 0.0023}  & {\bf 62.8}  & 0.054 \\
           & 1 &  2 & 10 &  2         & {\bf 2.66} & 40.7        &59.8             & {\bf 0.86}   &  0.093        &  0.0114          & 17.6         & 0.060\\
           & 1 &  2 &  5  &  1         & {\bf 1.50} &  --             & --                 & 0.64           &  0.089         &  0.0113         & 26.1         & 0.074 \\
           & 2 &  1 &10  & 2          & {\bf 1.71} & 36.7        & 64.5            & 0.56           & 0.111         &  0.0065           &263.9       & 0.067\\
           & 2 &  1 &  5  & {\bf 4}  & {\bf 2.15}  & 31.0        &{\bf 24.8}     & {\bf 0.75}   & 0.077        &  0.0069           & 28.3         &0.077\\
           & 2 &  2 &10  & 2          & 0.89         & 34.0         &154.1           & 0.65           & 0.289         &  0.0427         & 23.3        & 0.077\\
           & 2 &  2 &  5  &  2         & {\bf 1.23} & {\bf 52.6}  &72.2            &  {\bf 0.95}   & 0.196         &  0.0572        & 12.7        & 0.046 \\
           & 3 &  1 &10  & 2         & 0.63         & 38.4          &85.1             &  0.66           & 0.170         &  0.0211        & 0.6          & 0.071\\
           & 3 &  1 &  5  &  1        & 0.59         &  --               & --                 & 0.72           & 0.597          &  0.2090        & 21.5        & 0.074\\
           & 3 &  2 &10  &  2        & 0.67          & {\bf 46.6}  &140.0          & {\bf 0.86}    & 0.478         &  0.1316       & 22.6        & 0.082\\
           & 3 &  2 &  5  &  1        & 0.69          &  --              &--                   & 0.73            & 0.411         & 0.0916        & 20.9         & 0.089 \\
           & 5 &  1 & 10 &  1       & 0.33          & --                &--                  &  0.61            & 0.171          & 0.0159       & --             & 0.084 \\
           & 5 &  1 &  5  &  2       & 0.67          & 31.1           &49.6             &  0.63           & {\bf 0.061}   & 0.0144       & --            & 0.074 \\   
           & 5 &  2 & 10 &  1       & 0.15         &   --               &--                  &  0.72           & 0.234           & 0.0356        & --            & 0.090\\
           & 5 & 2  & 5   &  1        & 0.21        &   --               &--                  &  0.65            & 0.263           & 0.0377        & --            & 0.093 \\ \hline
CJSECC   & 1 & 1  & 10 & {\bf 4} & 3.72  & 27.0        &35.3            & 1.28             & {\bf 0.041}  & {\bf 0.0019}   & 33.4     & 0.004\\
           & 1 & 1  & 5   &  {\bf 3} & {\bf 2.74} & 32.8         &{\bf 31.7}    &  1.11            & 0.180          & 0.0192           & {\bf 111.9}  & 0.007\\  
           & 1 & 2  & 10 &  {\bf 3} & {\bf 2.00} & 42.6         &48.6           & 1.18              & 0.077          & 0.0076          & {\bf 50.5} & 0.011\\ 
           & 1 & 2  & 5   &  {\bf 3} & {\bf 1.96}  & {\bf 50.8} &56.5           & {\bf 0.95}     & {\bf 0.045}  & {\bf 0.0033}  & {\bf 71.0}   & 0.014\\    
           & 2 & 1  & 10 & {\bf 5}  & {\bf 2.27}  & 10.9         &{\bf 25.0}   & {\bf 0.87}     &  {\bf 0.057} &{\bf 0.0024}   & {\bf 106.3} & 0.014 \\ 
           & 2 & 1  & 5   & {\bf 4}  & {\bf 2.17}   &  20.7        &{\bf 33.0}   & 1.06            & {\bf 0.065}  & 0.0038          & {\bf 110.1} & 0.005 \\
           & 2 & 2  &10  & {\bf 4}  & {\bf 1.43}  & {\bf 49.9}   &{\bf 32.2}   & 1.13             & {\bf 0.053} & {\bf 0.0022}  &  --               & 0.015\\
           & 2 & 2  & 5   & {\bf 3}  & {\bf 1.46}  & {\bf 59.6}   &{\bf 25.0}   & {\bf 0.89}     & 0.089        &  0.0060          & 16.5           & 0.014 \\
          & 3  & 1  & 10 & {\bf 3}  & {\bf 1.11}  & {\bf 64.8}   &{\bf 34.5}   & 1.33              & {\bf 0.064} & 0.0219          & 180.4         & 0.006 \\
          & 3  & 1  & 5   & {\bf 3}   & {\bf 1.44} & 27.3           &65.4          & 1.12              & {\bf 0.076} & 0.0060          & {\bf 80.8}    & 0.013\\
          & 3  & 2  & 10 &  2         & 0.71        & {\bf 89.4}     &54.3          & {\bf 1.02}      & 0.139          & 0.0112        & --                 & 0.032\\
          & 3  & 2  & 5   &  2         & 0.95         & {\bf 85.6}    &105.0        & {\bf 0.93}     & 0.160         & 0.0132          & 49.2            & 0.032\\ 
          & 5  & 1  & 10 &  {\bf 3} & 0.62         & 12.4           &121.7        & {\bf 0.78}     & 0.202         & 0.0476          & --                 &  0.013 \\  
          & 5  & 1  & 5   &  {\bf 3} & 0.92         & {\bf 61.8}    &{\bf 31.6}   &  {\bf 0.87}   & {\bf 0.025} &{\bf 0.0007}    &  --                &  0.037 \\ 
          & 5  & 2  & 10 &   2        & 0.41        & {\bf 67.5}    &74.1           &   1.08          & 0.121         & 0.0190           & --                 &  0.027 \\ 
          & 5  & 2  & 5   &   2       & 0.50          &{\bf 110.4}  &57.9           & {\bf 1.02}  & {\bf 0.061} & 0.0038            &  --                 &  0.029\\ \hline
  SS   &     &     &       & 4 &  1.98               & 89.9          &  26.3            &  0.90         & 0.038          &  0.0018           & 50--150 & 0.048 \\ \hline          
\end{tabular}
\caption{The same as Table~1, but for the EEJS and CJSECC simulations}
\end{center}
\end{table}

\clearpage

\begin{figure}
\begin{center}
\includegraphics[width=0.8\textwidth]{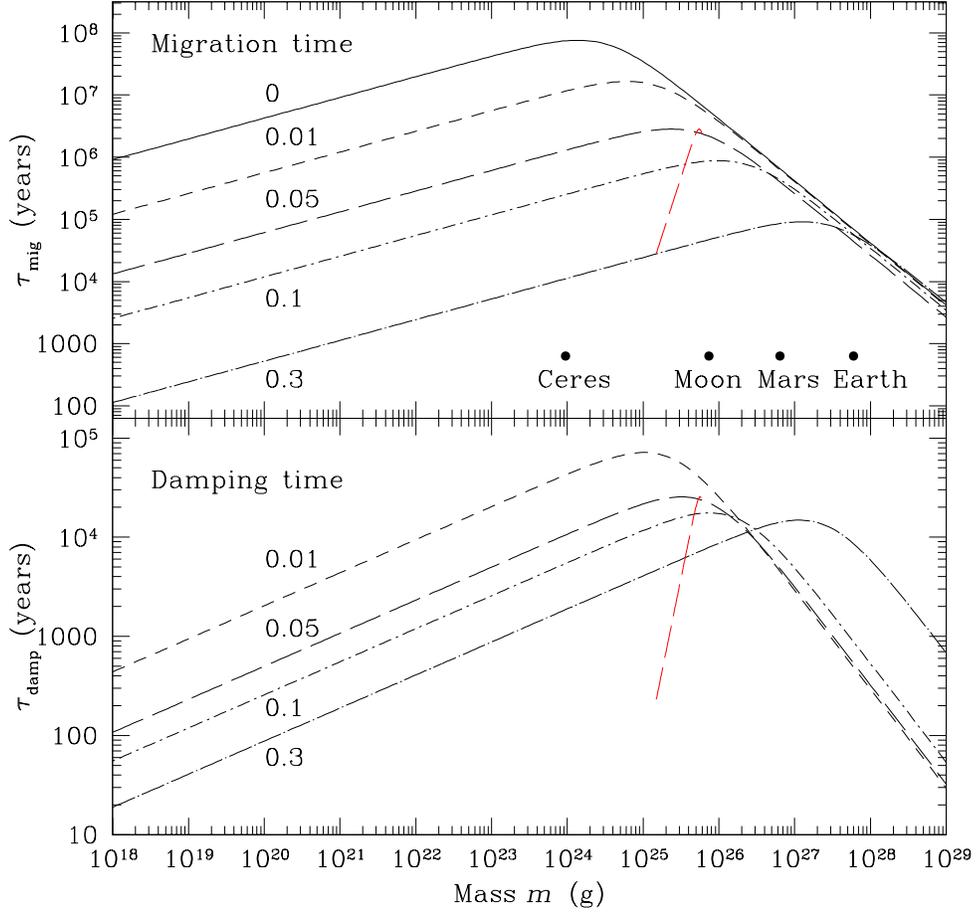}
\end{center}
\caption{Time scales for radial migration $\tau_{\rm mig}$ and damping of eccentricity $\tau_{\rm damp}$ 
due to gas drag and tidal interaction 
($\tau_{\rm mig} = a[(da/dt)_{\rm drag} + (da/dt)_{\rm tid}]^{-1}$
and $\tau_{\rm damp} = e[(de/dt)_{\rm drag} + (de/dt)_{\rm tid}]^{-1}$) at 1 AU with $\Sigma_{\rm gas} = 2000$ g cm$^{-2}$ and $i=0$.
The migration and damping rates due to gas drag, $(da/dt)_{\rm drag}$ and  $(de/dt)_{\rm drag}$, 
are derived after Adachi et al. (1976) with some corrections given in Inaba et al. (2001).
The labels to the lines represent $e$.
For $e=0.05$, we also plot the time scales with the enhancement of the gas drag force for $m_0 = 0.0025$ $M_{\oplus}$,
$m_{\rm 1} = 0.01$ $M_{\oplus}$, and $m_{\rm GI} = 10^{19}$ g (see Sec.~2.2.1). The masses of solar system bodies are also plotted in the upper panel.}
\end{figure}

\clearpage
\begin{figure}
\begin{center}
\includegraphics[width=0.85\textwidth]{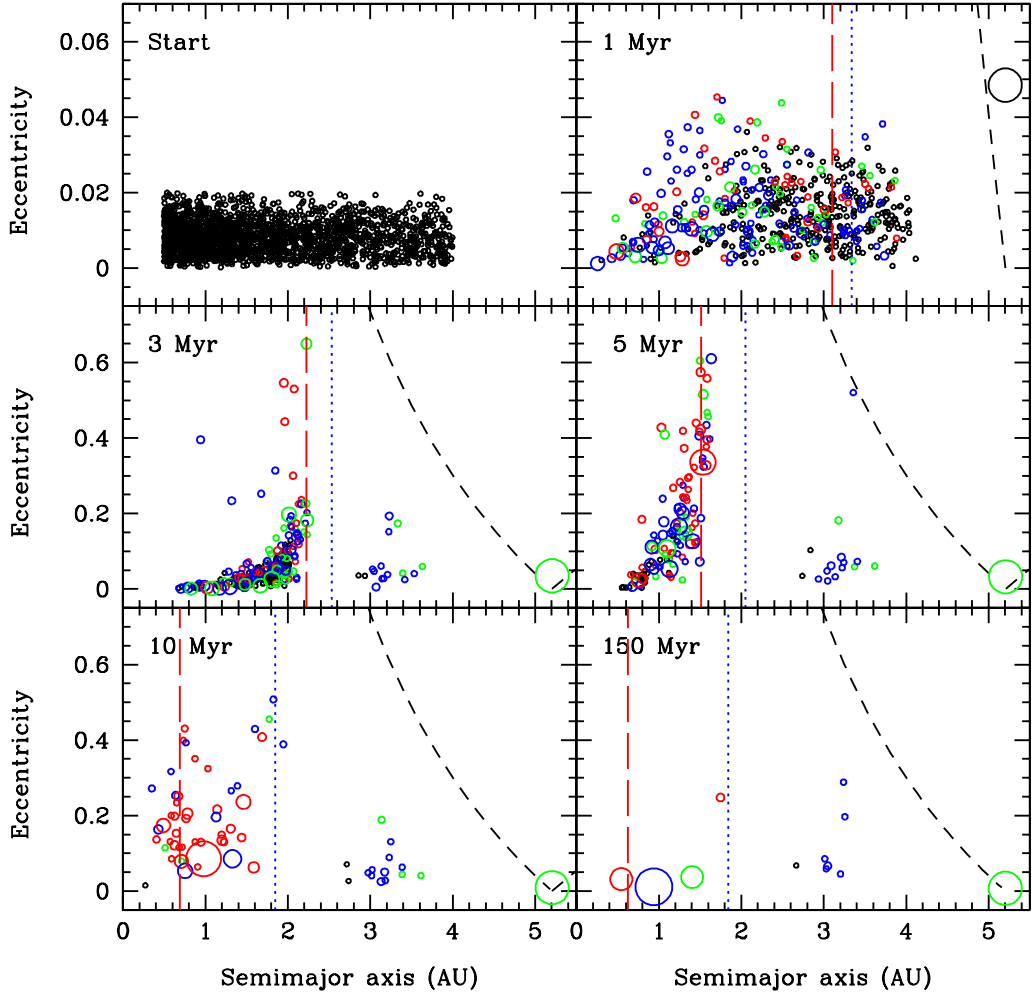}
\end{center}
\caption{Snapshots of an EJS simulation ($\tau_{\rm decay}$ = 1 Myr, $p = 2$, and $M_{\rm T}$ = 5 $M_{\oplus}$) 
on the $a-e$ plane. The size of each body is proportional to the radius, except Jupiter's size is modified to the 
Earth size on this plot.  The color represents spin rate $\omega$:  $\omega > \omega_{\rm crit}$,  
$\omega_{\rm crit} \ge \omega > 0.5$ $\omega_{\rm crit}$, 0.5 $\omega_{\rm crit} \ge \omega > 0$, and $\omega = 0$ for 
red, blue, green, and black bodies, respectively, where $\omega_{\rm crit}$ is the spin rate with which the centrifugal force 
balances with the gravity at the surface. The long-dashed and short-dashed lines are locations of the $\nu_5$ and $\nu_6$ resonances, respectively.
The black dashed curve is given by $a(1+e) = a_{\rm J}$;  particles in the right side of this curve experience orbital crossing with Jupiter.
Note the difference in scales of $e$ between the top panels and others. }  
\end{figure}

\clearpage
\begin{figure}
\begin{center}
\includegraphics[width=0.8\textwidth]{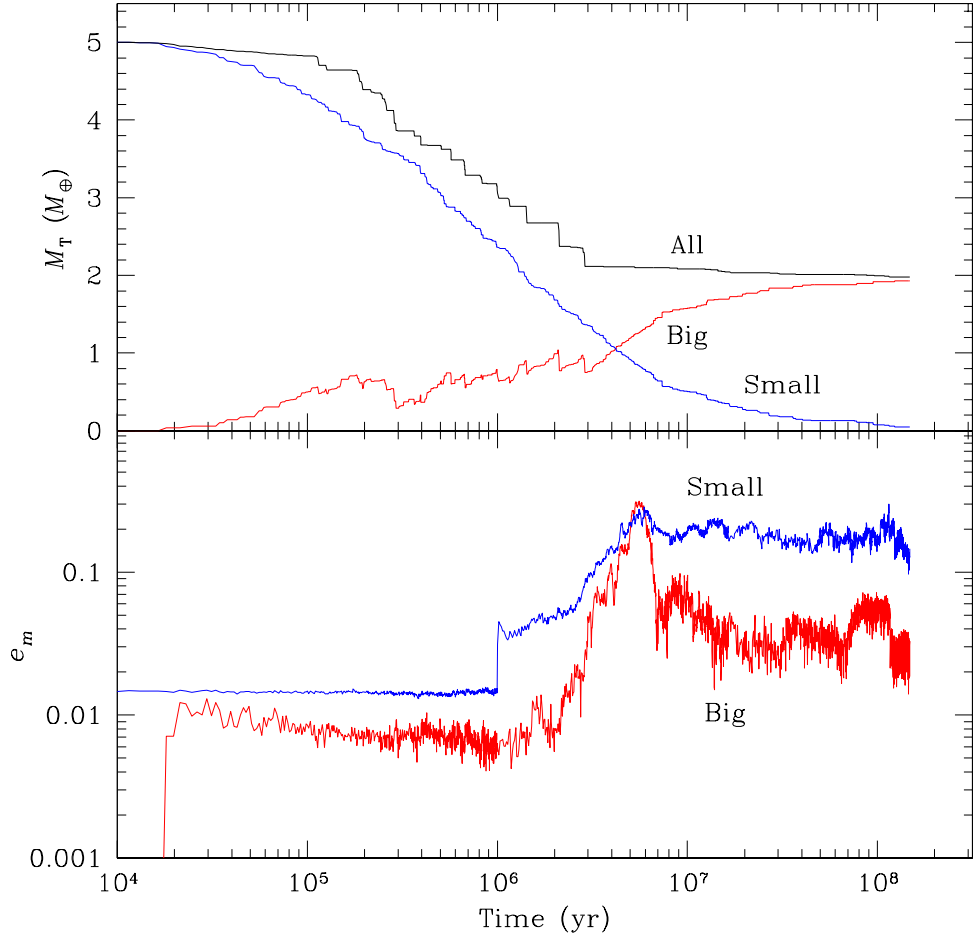}
\end{center}
\caption{Time evolutions of the total masses of small, big, and all particles (the upper panel) and 
the mass weighted mean eccentricities of small and big particles (the lower panel) for an EJS simulation.
We define mass of a small (big) particle to be below (above) $2 \times 10^{26}$ g. }
\end{figure}

\clearpage
\begin{figure}
\begin{center}
\includegraphics[width=0.8\textwidth]{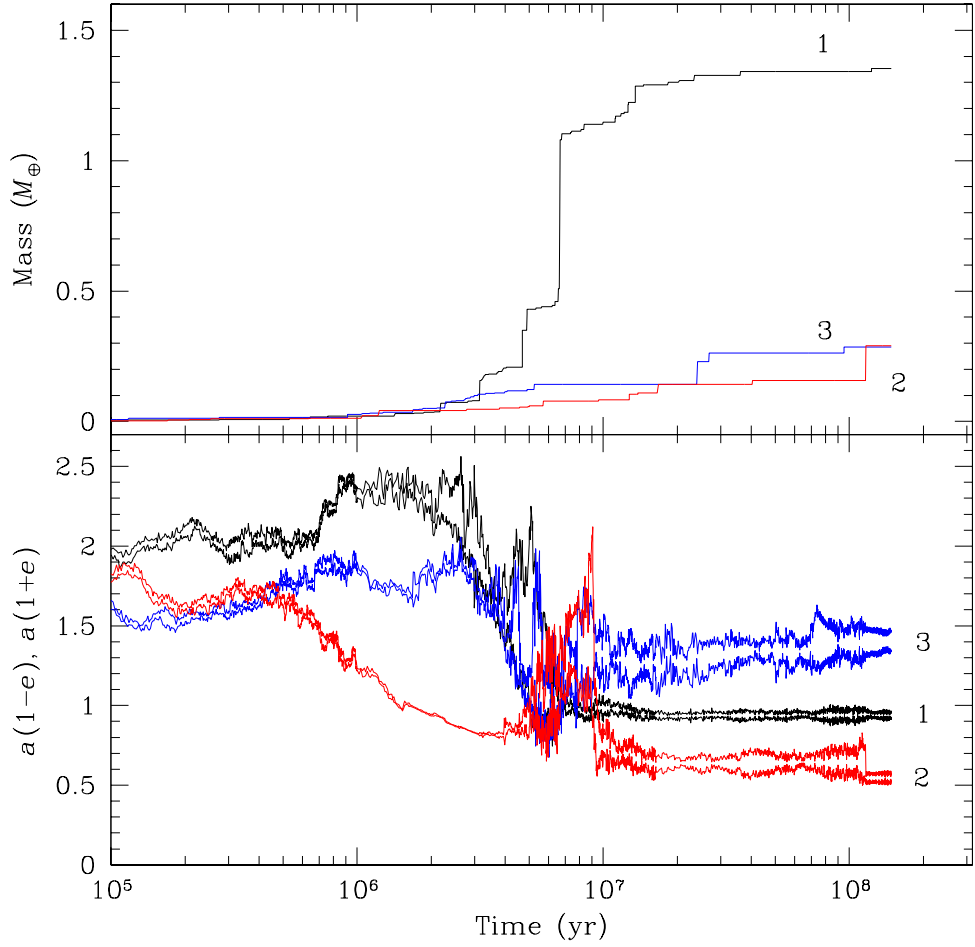}
\end{center}
\caption{The time evolutions of the masses (the upper panel) and 
the orbital excursions (the lower panel) of surviving planets in an EJS simulation.
The numbers of planets are given in the order of their masses in this figure.}
\end{figure}

\clearpage
\begin{figure}
\begin{center}
\includegraphics[width=0.85\textwidth]{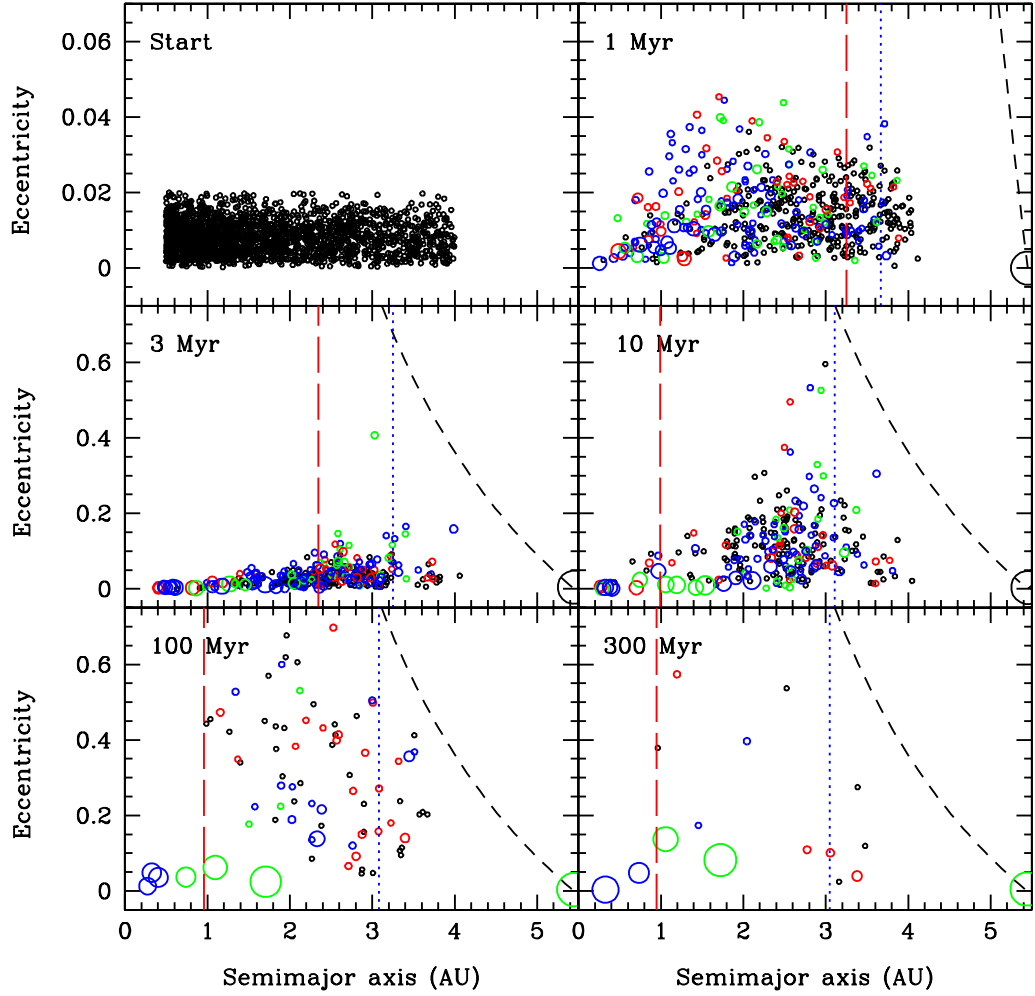}
\end{center}
\caption{The same as Fig.~2, but for a CJS simulation. Except for orbits of Jupiter and Saturn, other parameters are the same as those in Fig.~2.}

\end{figure}

\clearpage
\begin{figure}
\begin{center}
\includegraphics[width=0.8\textwidth]{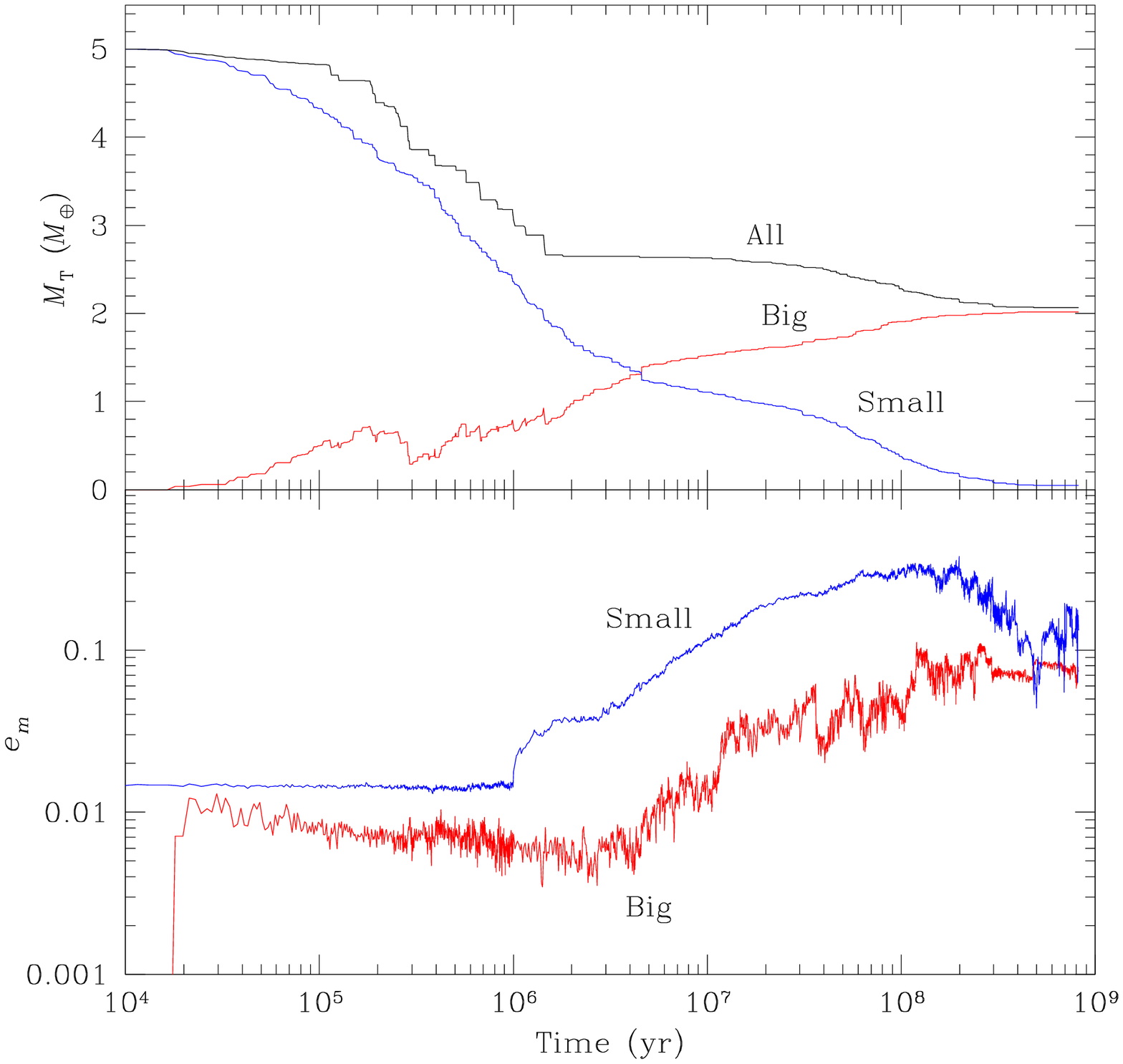}
\end{center}
\caption{The same as Fig.~3, but for a CJS simulation.}

\end{figure}

\clearpage
\begin{figure}
\begin{center}
\includegraphics[width=0.8\textwidth]{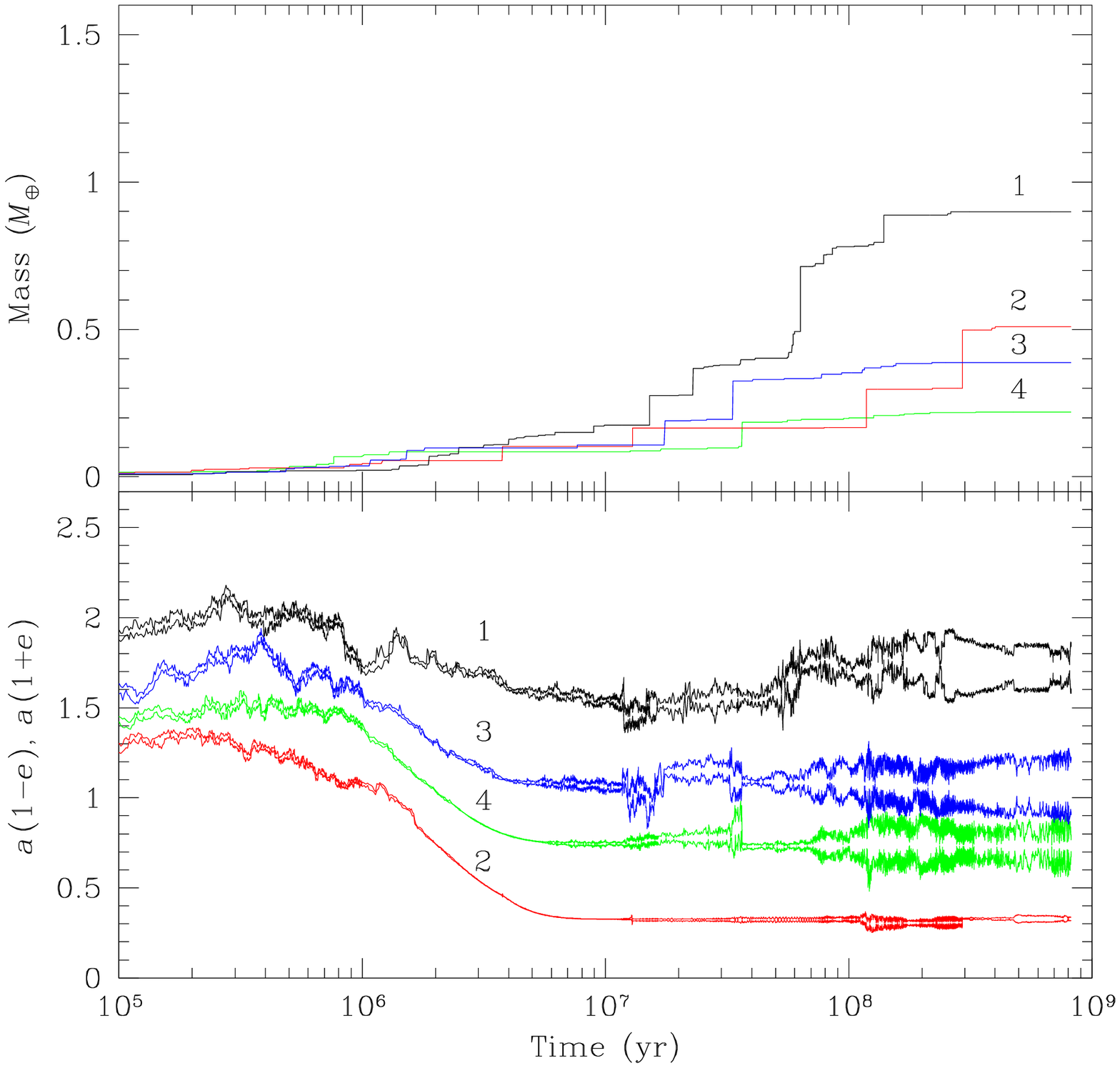}
\end{center}
\caption{The same as Fig.~4, but for a CJS simulation.}
\end{figure}

\clearpage
\begin{figure}
\begin{center}
\includegraphics[width=0.85\textwidth]{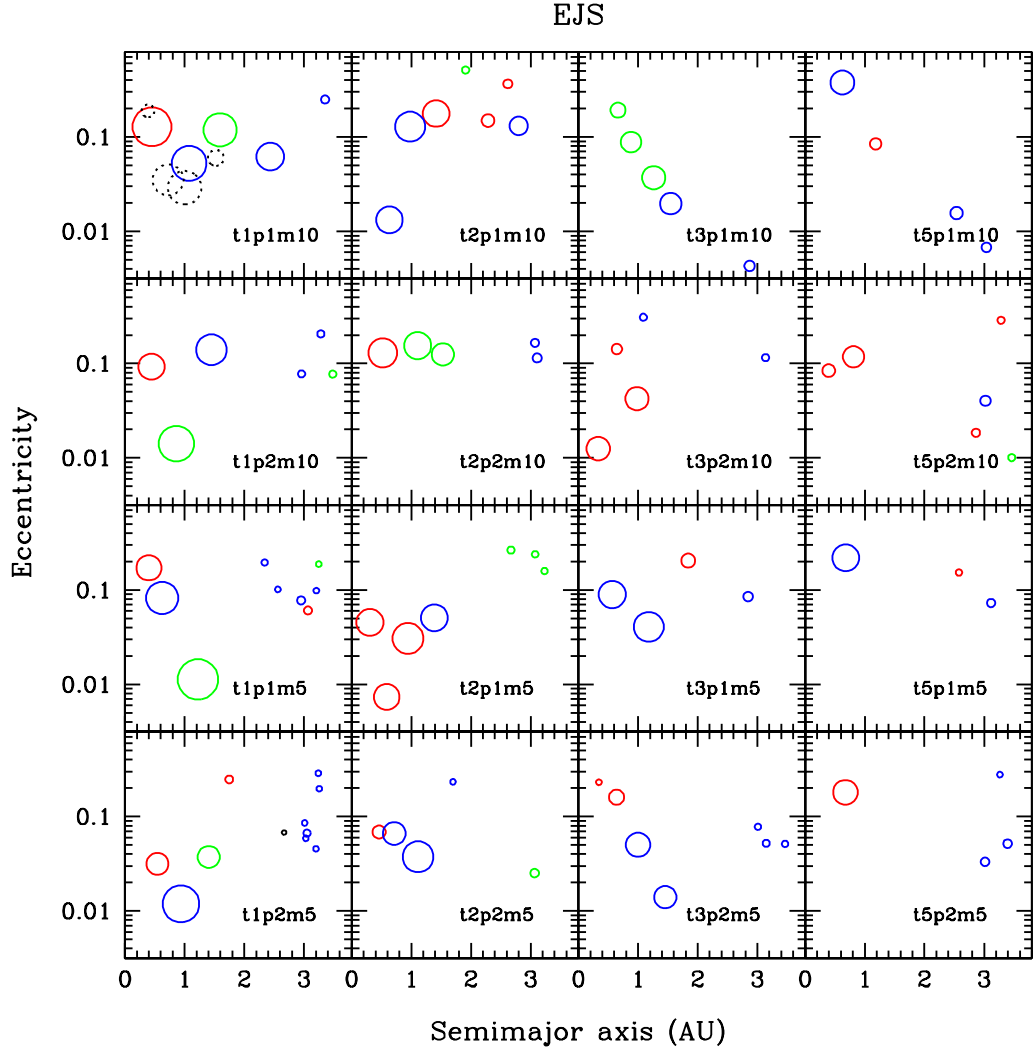}
\end{center}
\caption{Snapshots of all EJS simulations on the $a-e$ plane at the end of simulations. 
In each panel the numbers after $t$, $p$, 
and $m$ represent $\tau_{\rm decay}$ in units of Myr,  $p$ itself, and $M_{\rm T}/M_{\oplus}$. 
The colors of bodies represent the spin rates as well as Fig.~2.
In the panel at the upper left corner, we plot 
the current terrestrial planets with dashed circles.}

\end{figure}

\clearpage
\begin{figure}
\begin{center}
\includegraphics[width=0.85\textwidth]{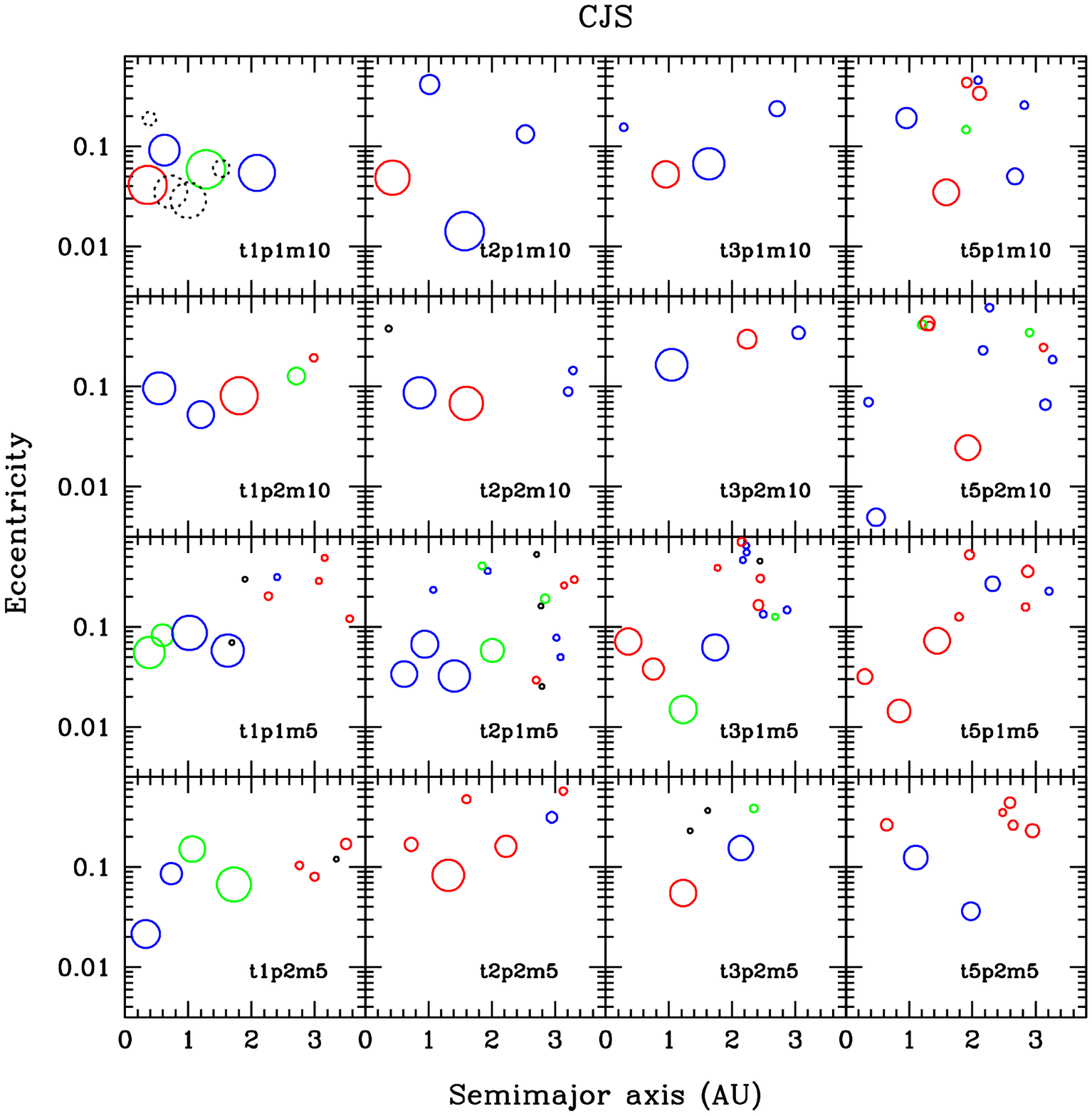}
\end{center}
\caption{The same as Fig.~8, but for the CJS simulations. }

\end{figure}

\clearpage
\begin{figure}
\begin{center}
\includegraphics[width=0.8\textwidth]{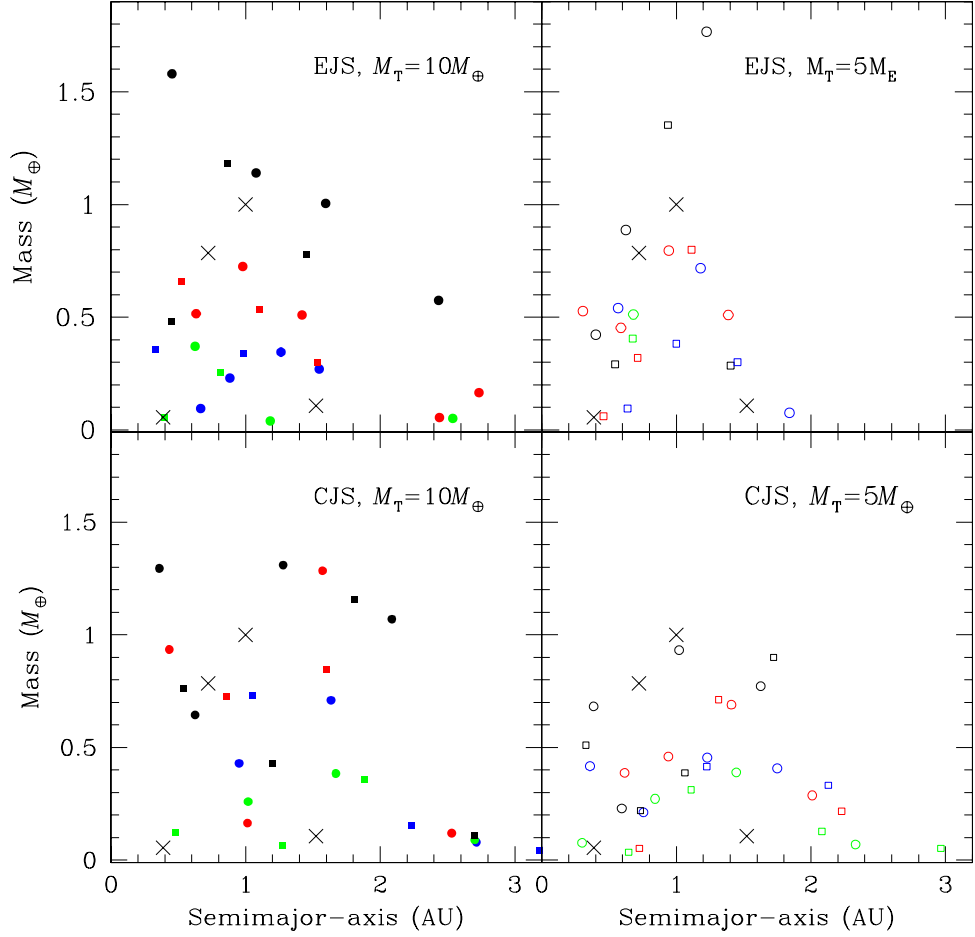}
\end{center}
\caption{Mass vs. semimajor axis for the EJS and CJS simulations. The filled and open symbol are for $M_{\rm T}/M_{\oplus} = 5$ and 10.
The color represents the gas decay time: $\tau_{\rm decay} = 1,2,3,$ and 5 Myr
for black, red, blue, and green symbols.  Circles and squares are for $p =1$ and 2. Cross marks represent the current terrestrial planets.}

\end{figure}

\clearpage
\begin{figure}
\begin{center}
\includegraphics[width=0.8\textwidth]{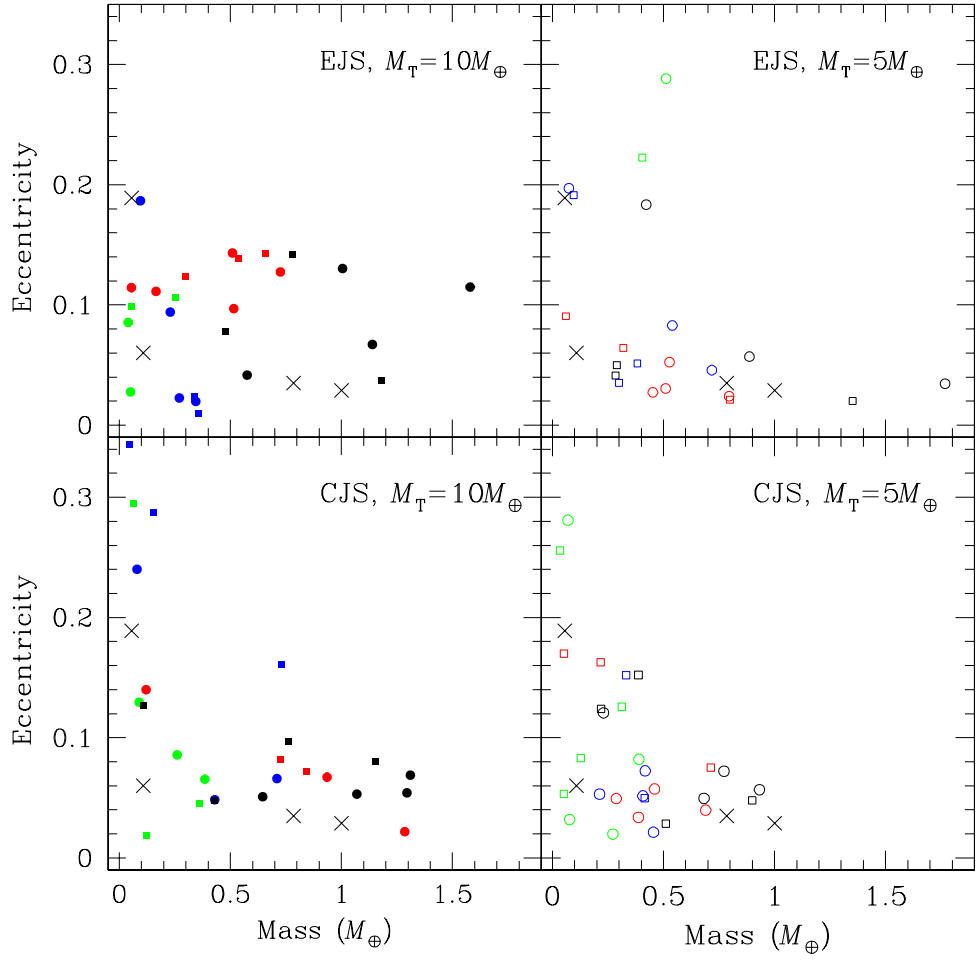}
\end{center}
\caption{Eccentricity vs. mass for the EJS and CJS simulations. See Fig.~10 for symbols.}

\end{figure}

\clearpage
\begin{figure}
\begin{center}
\includegraphics[width=0.8\textwidth]{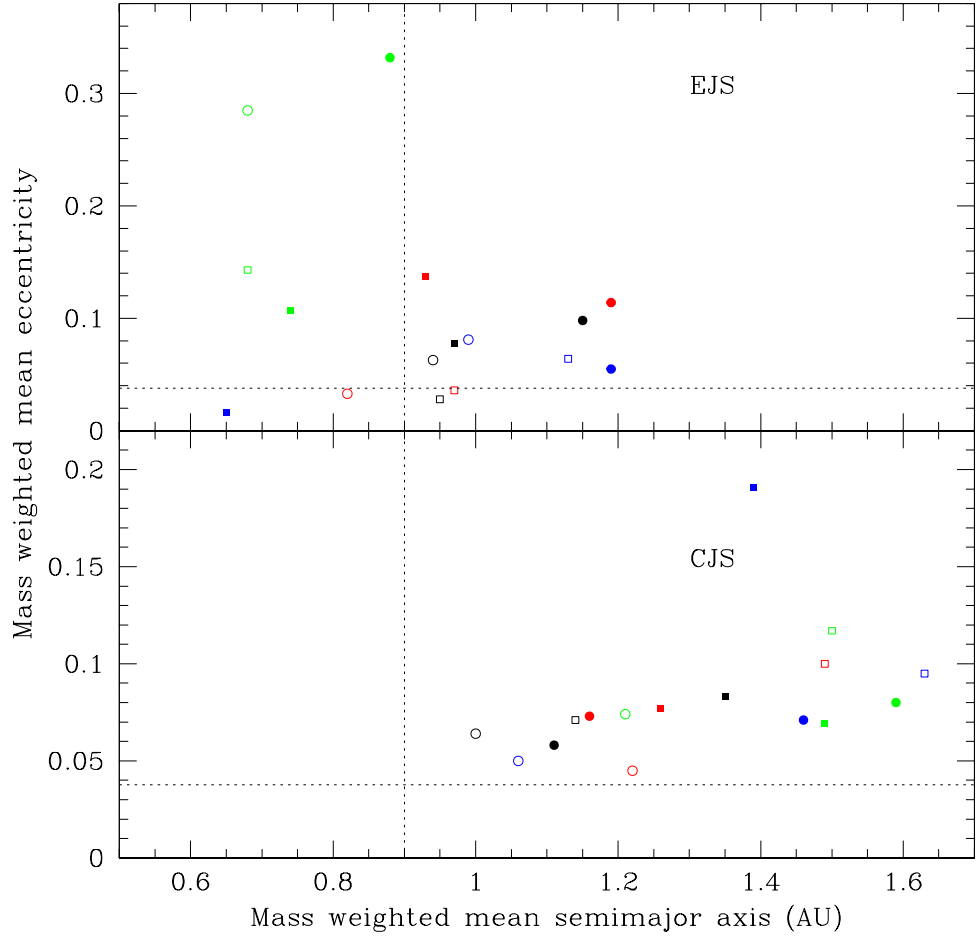}
\end{center}
\caption{Mass weighted mean eccentricity $e_m$ vs. mass weighted mean semimajor axis $a_m$ for the EJS and CJS simulations.
The dotted vertical and horizontal lines indicate the values for the current terrestrial planets. See Fig.~10 for symbols.}
\end{figure}

\clearpage
\begin{figure}
\begin{center}
\includegraphics[width=0.8\textwidth]{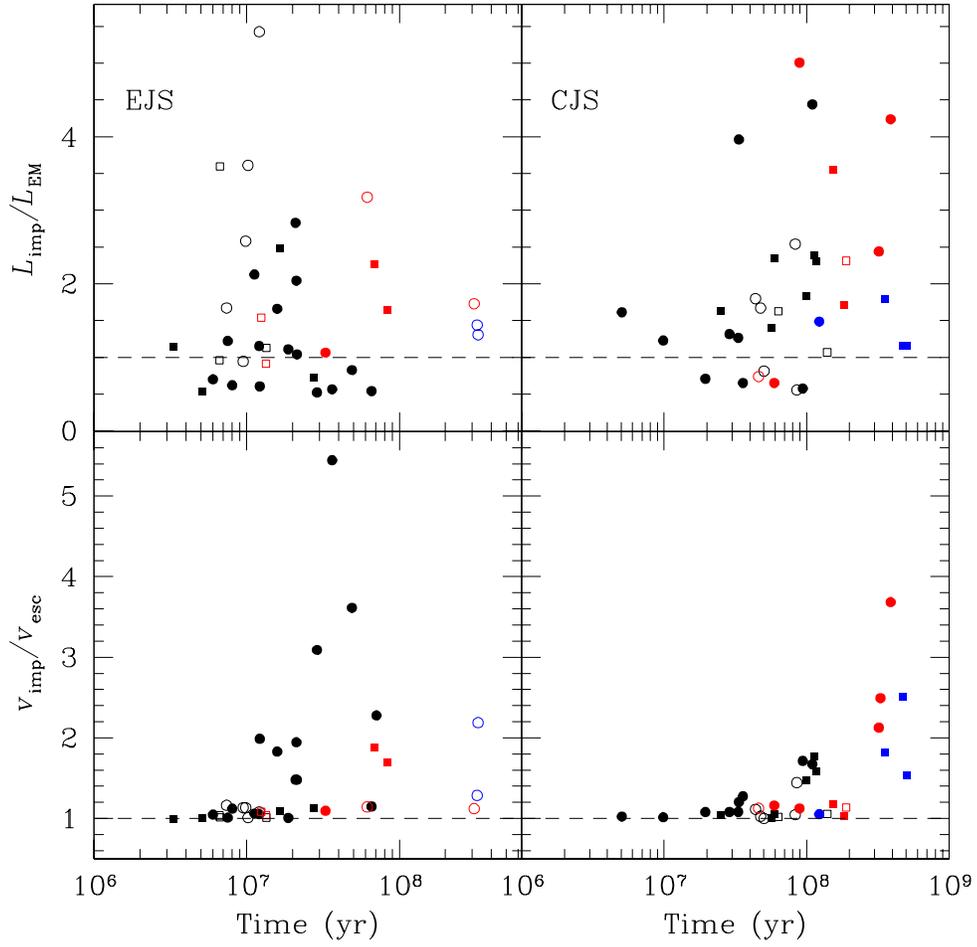}
\end{center}
\caption{Angular momenta and velocities of potentially Moon-forming impacts as functions of time for the EJS and CJS simulations.
See Fig.~10 for symbols.}

\end{figure}

\clearpage
\begin{figure}
\begin{center}
\includegraphics[width=0.85\textwidth]{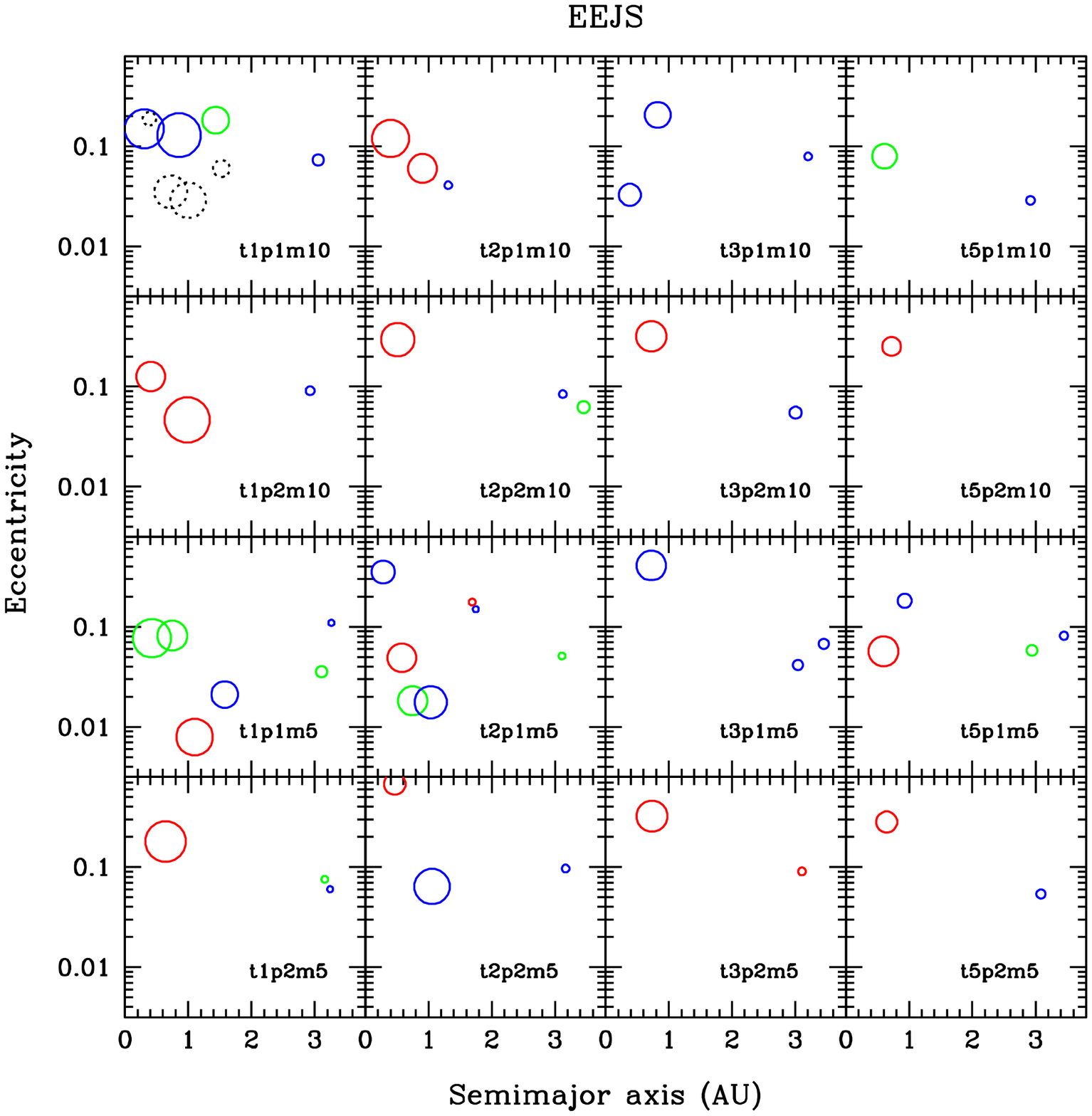}
\end{center}
\caption{The same as Fig.~8, but for the EEJS simulations.}

\end{figure}

\clearpage
\begin{figure}
\begin{center}
\includegraphics[width=0.85\textwidth]{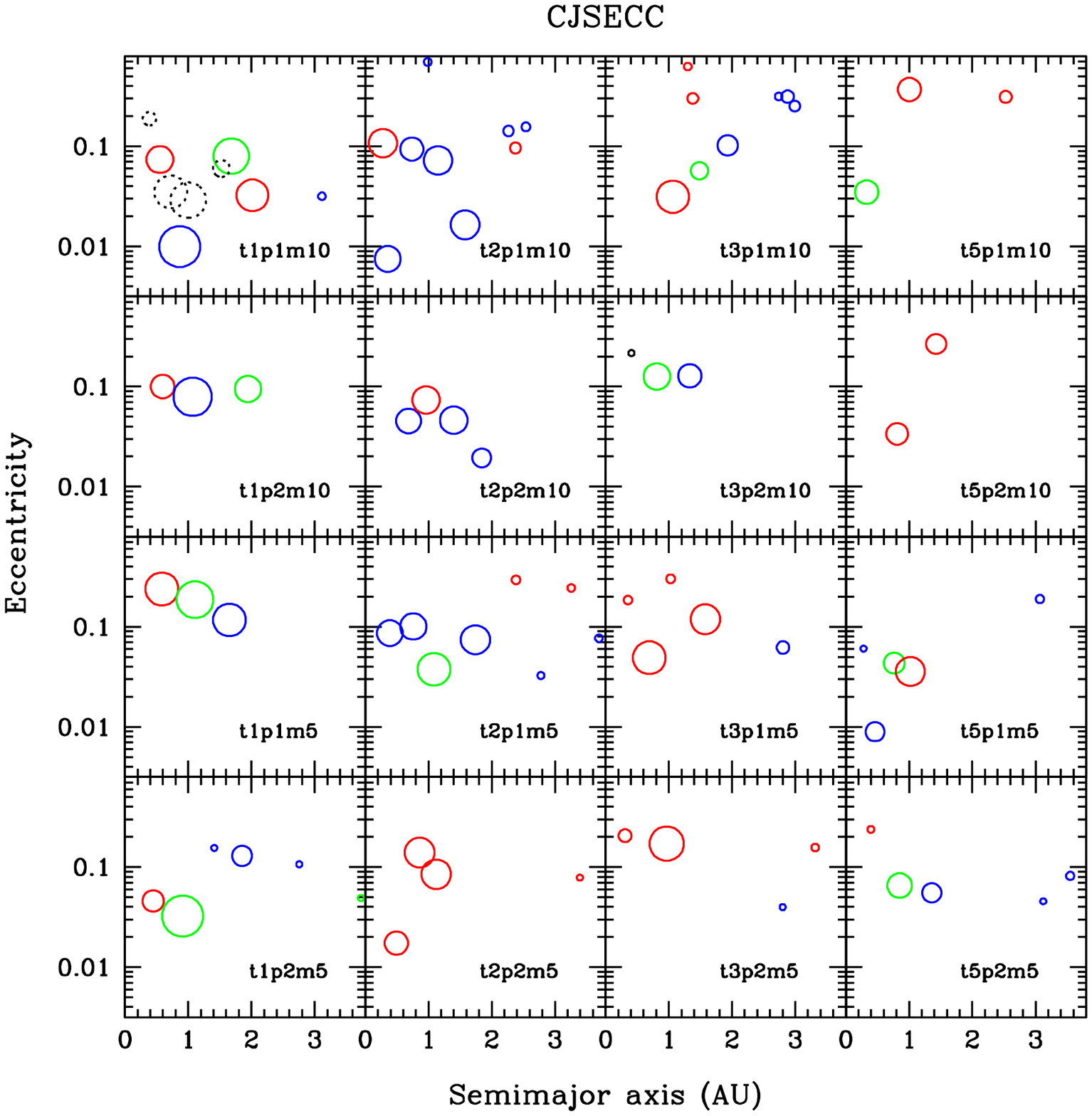}
\end{center}
\caption{The same as Fig.~8, but for the CJSECC simulations.} 

\end{figure}

\clearpage
\begin{figure}
\begin{center}
\includegraphics[width=0.8\textwidth]{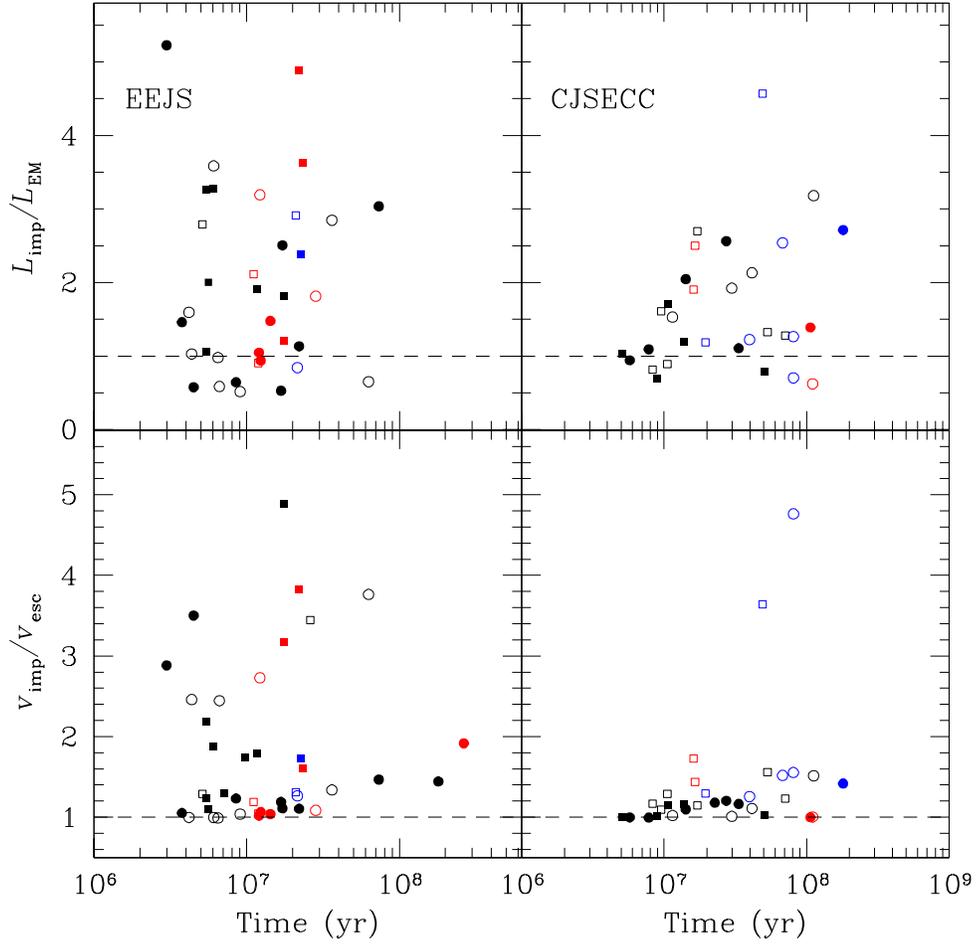}
\end{center}
\caption{The same as Fig.~13, but for the EEJS and CJSECC simulations.
Some impacts in the EEJS simulations have too large $L_{\rm imp}$ ($> 6$ $L_{\rm EM}$) and only their $v_{\rm imp}$'s are shown. } 

\end{figure}

\clearpage
\begin{figure}
\begin{center}
\includegraphics[width=0.85\textwidth]{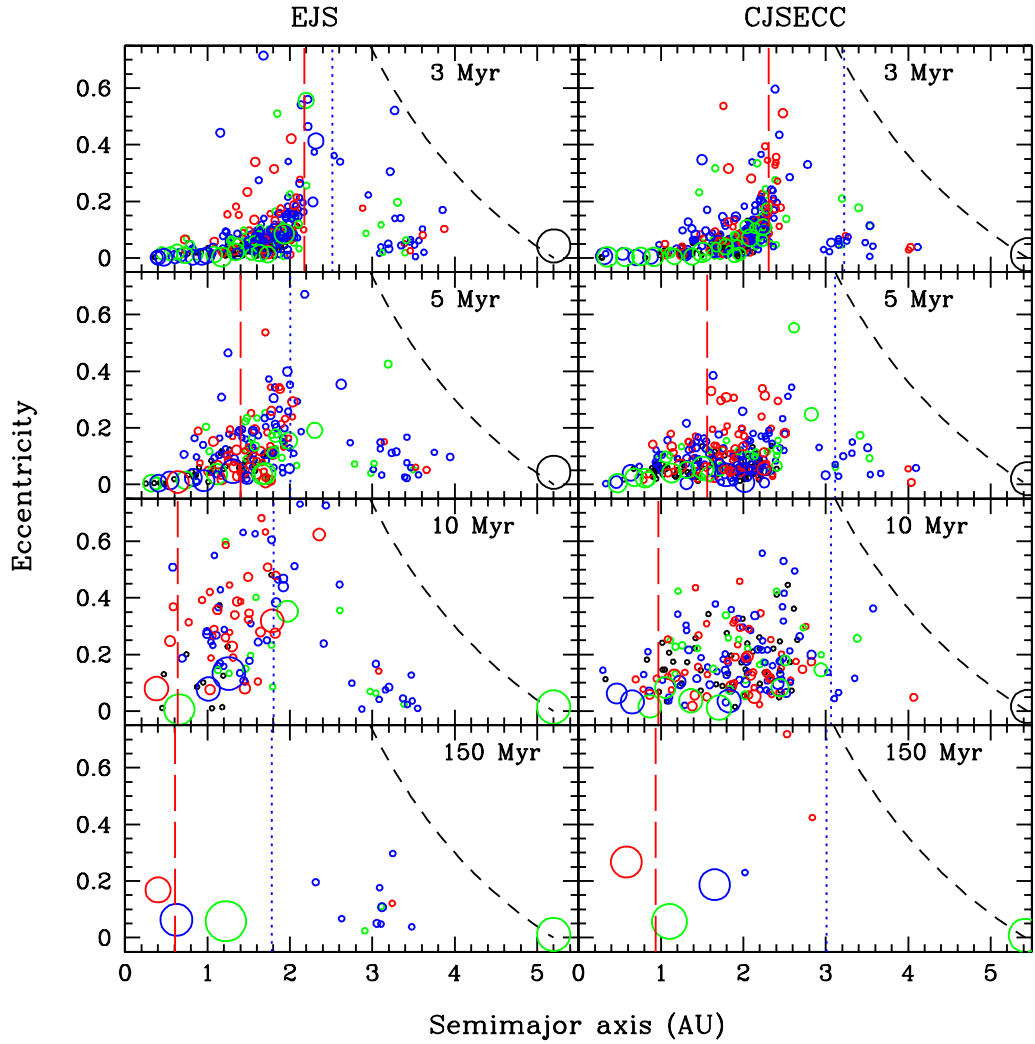}
\end{center}
\caption{Comparison between the EJS and CJSECC simulations. The same parameters are used for the two simulations except for 
orbits of Jupiter and Saturn: $\tau_{\rm decay} = 1$ Myr, $p =1$, and $M_{\rm T} = 5$ $M_{\oplus}$. The format for colors and lines is after Fig.~2.}

\end{figure}

\clearpage
\begin{figure}
\begin{center}
\includegraphics[width=0.85\textwidth]{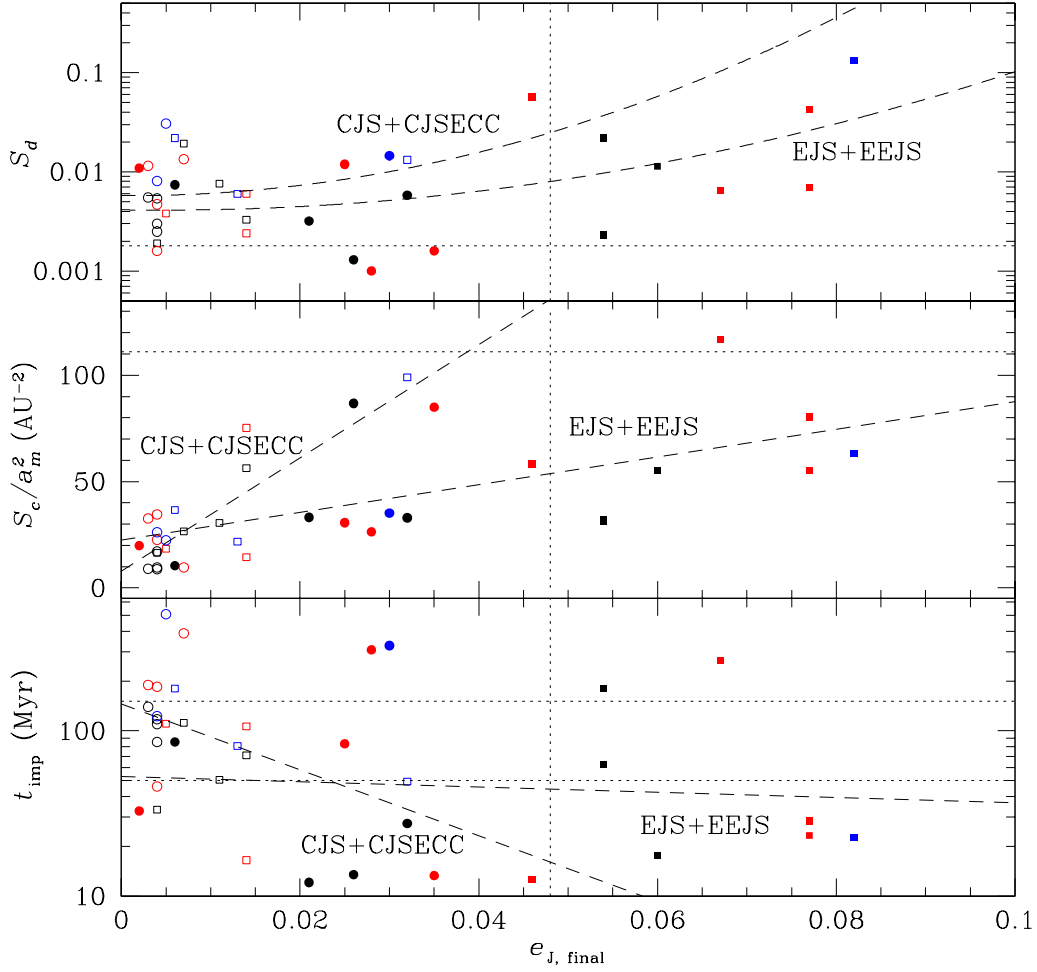}
\end{center}
\caption{Angular momentum deficit $S_d$ (upper panel), radial mass concentration parameter $S_c/a_m^2$ (middle panel), 
and time of the last one of potentially Moon-forming impacts $t_{\rm imp}$ (lower panel) vs. final eccentricity of Jupiter $e_{\rm J, final}$.
Filled circles, filled squares, open circles, and open squares are results from the EJS, EEJS, CJS, and CJSECC simulations, respectively.
The color represents the gas decay time: $\tau_{\rm decay} = 1,2,$ and 3 Myr for black, red, and blue symbols. 
A fit to a quartic function is applied to $\log{S_d}$, while linear fits are applied to $S_c/a_m^2$ and $\log{t_{\rm imp}}$ for sets of 
the EJS+EEJS and CJS+CJSECC simulations, respectively. Horizontal and vertical dotted lines indicate the current values (for $t_{\rm imp}$, 
a possible range indicated from isotope records is between two dotted lines)}

\end{figure}

\clearpage
\begin{figure}
\begin{center}
\includegraphics[width=0.85\textwidth]{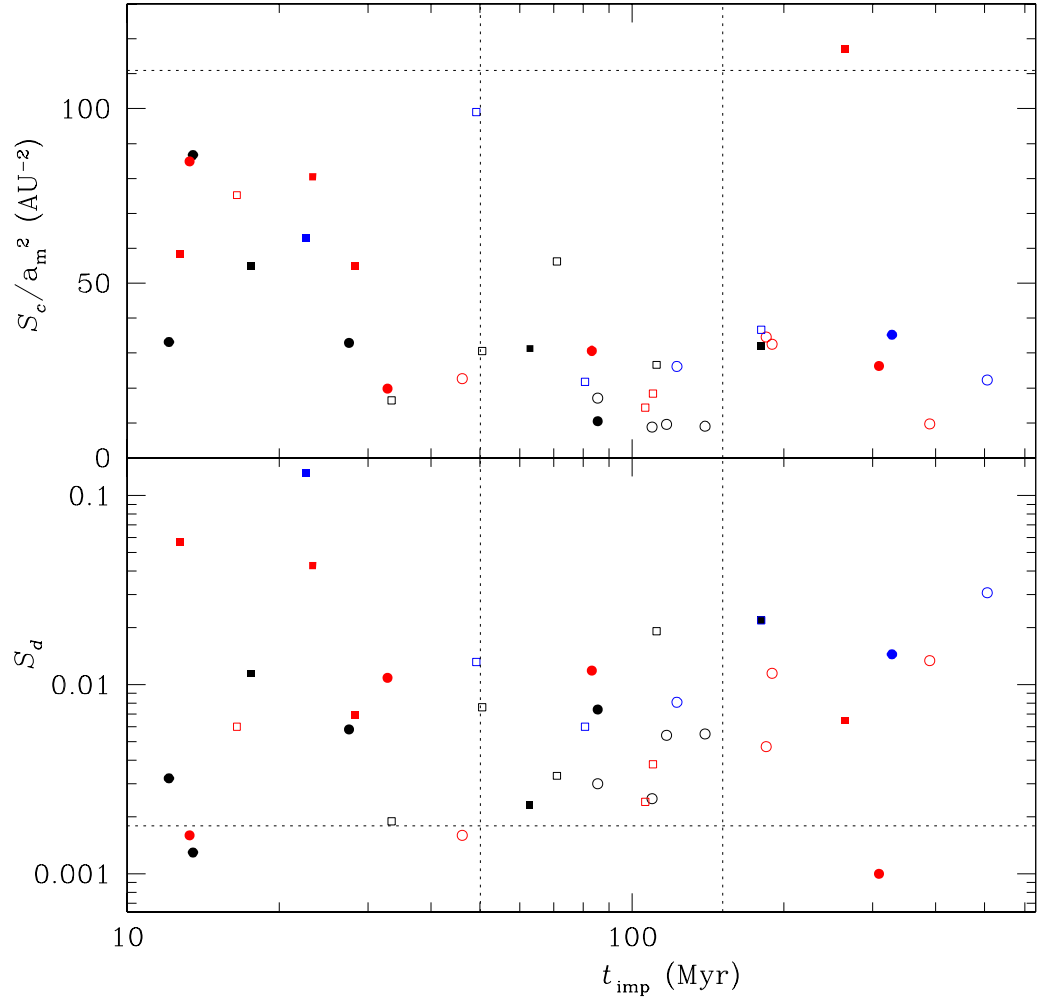}
\end{center}
\caption{Radial mass concentration parameter $S_c/a_m^2$ (upper panel) and angular momentum deficit $S_d$ (lower panel) vs. 
time of the Moon-forming impact $t_{\rm imp}$. The data are the same as those in Fig.~18.}

\end{figure}

\end{document}